\documentclass[fleqn,usenatbib]{mnras}

\usepackage{newtxtext,newtxmath}

\usepackage[T1]{fontenc}

\DeclareRobustCommand{\VAN}[3]{#2}
\let\VANthebibliography\thebibliography
\def\thebibliography{\DeclareRobustCommand{\VAN}[3]{##3}\VANthebibliography}

\usepackage{graphicx}	
\usepackage{amsmath}	

\title[Protostellar Disk Formation]{Protostellar Disks Fed By Dense Collapsing Gravo-Magneto-Sheetlets} 

\author[Y. Tu et al.]{
Yisheng Tu,$^{1}$\thanks{E-mail: yt2cr@virginia.edu}
Zhi-Yun Li,$^{1}$
Ka Ho Lam$^{1}$
Kengo Tomida$^{2}$
Chun-Yen Hsu$^{1}$
\\
$^{1}$Astronomy Department, University of Virginia, Charlottesville, VA 22904, USA\\
$^{2}$Tohoku University Astronomical Institute 6-3, Aramaki Aza-Aoba, Aoba-ku. Sendai, Miyagi 980-8578
}

\date{Accepted XXX. Received YYY; in original form ZZZ}

\pubyear{2015}

\begin{document}
\label{firstpage}
\pagerange{\pageref{firstpage}--\pageref{lastpage}}
\maketitle

\begin{abstract} 
Stars form from the gravitational collapse of turbulent, magnetized molecular cloud cores. Our non-ideal MHD simulations reveal that the intrinsically anisotropic magnetic resistance to gravity during the core collapse naturally generates dense gravo-magneto-sheetlets within inner protostellar envelopes -- disrupted versions of classical sheet-like pseudodisks. They are embedded in a magnetically dominant background, where less dense materials flow along the local magnetic field lines and accumulate in the dense sheetlets. The sheetlets, which feed the disk predominantly through its upper and lower surfaces, are the primary channels for mass and angular momentum transfer from the envelope to the disk. The protostellar disk inherits a small fraction (up to 10\%) of the magnetic flux from the envelope, resulting in a disk-averaged net vertical field strength of 1-10 mG and a somewhat stronger toroidal field, potentially detectable through ALMA Zeeman observations. The inherited magnetic field from the envelope plays a dominant role in disk angular momentum evolution, enabling the formation of gravitationally stable disks in cases where the disk field is relatively well-coupled to the gas. Its influence remains significant even in marginally gravitationally unstable disks formed in the more magnetically diffusive cases, removing angular momentum at a rate comparable to or greater than that caused by spiral arms. The magnetically driven disk evolution is consistent with the apparent scarcity of prominent spirals capable of driving rapid accretion in deeply embedded protostellar disks. The dense gravo-magneto-sheetlets observed in our simulations may correspond to the ``accretion streamers" increasingly detected around protostars.

\end{abstract}




\section{Introduction}
\label{sec:intro}

Circumstellar disks play a central role in the formation of both stars and planets. How they form and evolve has been a topic of active research for decades. Much of the recent effort has been focused on the role of magnetic fields in disk formation, as reviewed in, e.g. \citet{Li2014PPVI, Wurster2018, ZhaoTomida2020, Tsukamoto2022PPVII}. There appears to be a general agreement that the magnetic field can strongly modify but does not prevent the formation of a rotationally supported disk in the presence of a realistic level of magnetic diffusion. With the question of whether disks can form in magnetized molecular cloud cores largely settled, it is natural to shift the focus to characterizing the properties of the formed protostellar disks and their connection with the collapsing envelope feeding them. 

The protostellar disk's magnetic field strength is particularly important to quantify. It determines not only the structure and evolution of the disk during the protostellar phase but also sets the initial conditions for the subsequent disk evolution in the protoplanetary phase. For example, if little or no magnetic flux is inherited from the envelope, the disk would have to rely on gravitational torque to transport angular momentum, with its structure and evolution dominated by spiral arms rather than magneto-rotation instability or magnetically driven winds, as recently reviewed by \citet{Lesur2022PPVII}. These authors emphasized: ``A major uncertainty here is the field strength, which is not predicted by any complete theory,
and for which we have few empirical constraints." We fully agree with this assessment. Some empirical constraints on the field strength on the disk scale have recently been obtained with ALMA CN Zeeman observations, which provided upper limits of several to tens of milli-Gausses \citep{Vlemmings2019, Harrison2021}. Our investigation aims to take initial steps toward a complete theory capable of predicting the disk field strength and confronting the prediction with observations. 

Another goal is to quantify the role of the magnetic field in structuring the inner protostellar envelope. It is timely because ALMA has been observing the inner envelope in detail in molecular lines and, increasingly, dust continuum polarization \citep[e.g.][]{Pattle2022}; the latter trace magnetic fields on the envelope scale. 
It is natural to ask how the observationally accessible magnetic field affects the envelope's structure and dynamics and how the magnetic structuring of the envelope affects the disk formation. 

There is little doubt that a dynamically significant magnetic field can strongly affect the mass distribution of the collapsing protostellar envelope. It is well known that magnetic fields tend to deflect the gravity-induced collapsing motions toward a direction along the field lines through a tension force, forming a dense gravo-magneto-sheet near the magnetic equator -- the classical ``pseudodisk" \citep{Galli&Shu1993}. Its formation is an unavoidable consequence of the anisotropic nature of the magnetic resistance to gravity. A realistic level of turbulence is expected to warp the dense sheet out of the equatorial plane and potentially break it up into smaller pieces, forming a highly perturbed three-dimensional structure that we will term ``gravo-magneto-sheetlets" (or ``sheetlets" for short). They show up as filaments in the column density maps of many disk formation simulations that include both magnetic fields and turbulence (\citealp[e.g.][]{Santos-Lima2012, Seifried2013, Li2014, Seifried2015_turb, Kuffmeier_2017, Matsumoto2017, Gray2018, Lam2019, Wurster2020_I, Wurster2020_II}; see \citealp{Tsukamoto2022PPVII} for a review), although the deep connection between the apparent filaments in the column density map and the classical pseudodisk of \citet{Galli&Shu1993} was rarely recognized. We aim to build a conceptual framework of protostellar envelopes dominated by gravo-magneto-sheetlets by examining such dense thin structures more closely than before, focusing on their contributions to the envelope's mass and angular momentum budgets, their effects on disk formation, and their connection to observations, particularly accretion streamers that are increasingly detected around Class 0 \citep[e.g.][]{Hsieh2020, Pineda2020, Murillo2022, Thieme2022} and Class I protostars \citep[e.g.][]{Yen2014, Yen2019, Valdivia-Mena2022}. 


The paper is organized as follows. Section~\ref{sec:problem_setup} describes the numerical setup of the problem, while  section~\ref{sec:simulation_result_overview} overviews our simulation results and defines terms used in our analysis. We focus on the formation of gravo-magneto-sheetlets and their leading roles in the structure and dynamics of the inner protostellar envelope in \S~\ref{sec:envelope}. It is followed by a discussion of the transition between the envelope and disk in \S~\ref{sec:Envelope2Disk}, emphasizing the magnetic flux fraction the disk inherits from the envelope. Section~\ref{sec:DiskOnly} discusses the properties of the formed disks, focusing on the role of the magnetic field in transporting angular momentum. We discuss our main results in \S~\ref{sec:discussion} and conclude in \S~\ref{sec:conclusion}.

\section{Problem Setup}
\label{sec:problem_setup}
%
%

The formation of protostellar disks from the collapse of dense magnetized molecular cloud cores is a complex process involving non-ideal MHD effects (particularly ambipolar diffusion) and turbulence. The following set of equations governs this turbulent, non-ideal MHD process:
\begin{equation}
    \frac{\partial\rho}{\partial t} + \nabla\cdot(\rho\mathbfit{v}) = 0,
\end{equation}
\begin{equation}
    \rho\frac{\partial \mathbfit{u}}{\partial t} + \rho(\mathbfit{u}\cdot\nabla)\mathbfit{u} = -\nabla P + \frac{1}{c}\mathbfit{J}\times\mathbfit{B} - \rho\nabla\Phi_g,
    \label{equ:mhd momentum equation}
\end{equation}
\begin{equation}
    \frac{\partial \mathbfit{B}}{\partial t} = \nabla\times(\mathbfit{v}\times\mathbfit{B}) - \frac{4\pi}{c}\nabla\times(\eta_A\mathbfit{J}_\perp),
\end{equation}
\begin{equation}
    \nabla^2\Phi_g = 4\pi G\rho
\end{equation}
where $\mathbfit{J} = (c/4\pi)\nabla\times\mathbfit{B}$ is the current density, $\mathbfit{J}_\perp = [(\mathbfit{J}\times\mathbfit{B})\times\mathbfit{B}] / B^2$ the component of current density perpendicular to the magnetic field, and $\eta_A$ the ambipolar diffusivity. Other symbols have their usual meanings.  

To simplify calculations, we adopted an equation of state consisting of four power-laws in four different density regimes based on the radiation magnetohydrodynamic calculations of \citet{Tomida2013}. 
The transition densities between the power-laws are chosen as $\rho_1 = 10^{-13}\mathrm{g\ cm^{-3}}$, $ \rho_2=3.16\times10^{-12}\mathrm{g\ cm^{-3}}$ and $\rho_3=5.66\times10^{-9}\mathrm{g\ cm^{-3}}$, with corresponding adiabatic constant $\gamma_1 = 5/3$, $\gamma_2=1.4$, and $\gamma_3=1.1$, respectively. 
At low densities $\rho < \rho_1$, we assume the gas is isothermal with $\gamma_0 = 1$.

Our simulation setup closely follows that of \cite{Lam2019}, except for our use of Adaptive Mesh Refinement (AMR), which is required to better resolve the formed disks, a key goal of our investigation. The AMR capability is afforded by the \textsc{ATHENA++} code \citep{Stone2008, Stone2020}, which includes a newly developed self-gravity solver specifically designed for AMR \citep{Tomida2023}. We used the full multigrid (FMG) mode of the multigrid self-gravity solver and obtained the boundary conditions for the gravitation potential using multipole expansion \citep{Tomida2023}. The governing equations are solved in a Cartesian coordinate system, with standard outflow boundary conditions implemented at all simulation domain boundaries.  

We used four strategies to speed up the simulations so that they could be completed in a reasonable amount of time. Firstly, we use the AMR capability to concentrate computational resources around the central star region. A refinement level is added when the local cell size exceeds 1/16 of the local Jeans length. Secondly, we adopt the sink particle technique of \citet{Lam2019} to treat protostellar accretion. A sink particle is created when the grid reaches the highest refinement level, and the local cell size exceeds 1/16 of the local Jeans length. After creation on the finest grid, the sink particle accretes mass and the radial component of the linear momentum from a sink region of $3\times 3\times 3$ cells, leaving the magnetic field untouched in the sink region. Thirdly, we cap the ambipolar diffusion using equation~(7) of \citet{Lam2019} to avoid 
prohibitively small time steps. This treatment primarily affects very low-density regions, and we monitor the simulation to make sure that its effect is minimal. Finally, we cap the Alfv\'en speed by adding small amounts of mass in highly evacuated cells and monitor the total added mass to ensure that it stays negligible compared to the mass in the computation domain. With these strategies, the models presented in this work each cost between 50,000 to 100,000 CPU core hours.
%

The simulation is initialized by putting a pseudo-Bonner-Ebert sphere of mass 0.5~$M_\odot$ and radius 2000 au at the center of the simulation domain with a box size of 10,000~au on each side. The density profile of the pesudo-Bonner-Ebert sphere is given by
\begin{equation}
    \rho(r) = \frac{\rho_0}{1 + (r/r_c)^2}
\end{equation}
where $\rho_0=4.6\times10^{-17}~\mathrm{g \ cm^{-3}}$ is the central density, and $r_c\approx670~\mathrm{au}$ is a radius characterizing the size of the central region of relatively uniform density distribution. We fill the space outside the sphere with diffuse material of a low density $4.56\times10^{-20}\ \mathrm{g\ cm^{-3}}$. The total mass in the simulation domain is $0.56M_\odot$, enough to form a star of a few tenths of the solar mass that is typical of low-mass stars \citep[e.g.][]{Offner2014PPVI}. We assume the core has an initial solid body rotation rate of $6.16\times10^{-13}\ \mathrm{s^{-1}}$ along the $\hat{z}$ axis and an initially uniform magnetic field of $2.2\times10^{-4}$ G (corresponding to a dimensionless mass-to-flux ratio of $2.6$ in units of the critical value $1/[2\pi G^{1/2}]$), also along the $\hat{z}$ axis. 

As in \cite{Lam2019}, we initialize the simulation with a turbulent velocity field of an $k^{-2}$ power spectrum \citep{Kolmogorov1941, Gong2011}. The net angular momentum from the turbulence is removed from the initial state. The velocity field's amplitude is normalized to obtain a root-mean-square Mach number $\mathcal{M}=1$ for most of our runs. We also consider two cases with, respectively, weaker turbulence of $\mathcal{M}=0.5$ and no turbulence ($\mathcal{M}=0$). The parameters of all models considered in this paper are summarized in Table~\ref{tab:para}.

\begin{table}
    \centering
    \begin{tabular}{l c c l}
        Model name & $\mathcal{M}$ & $\eta_0 / \eta_R$ & Comments \\
        \hline
        M1.0AD1.0 & 1.0 & 1.0 & Reference model \\
        M0.5AD1.0 & 0.5 & 1.0 & Weaker turbulence model \\
        M0.0AD1.0 & 0.0 & 1.0 & Non-turbulent model \\
        M1.0AD2.0 & 1.0 & 2.0 & More diffusive model \\
        M1.0AD10. & 1.0 & 10. & Most diffusive model \\
        M1.0AD0.1 & 1.0 & 0.1 & Least diffusive model \\
        \hline
    \end{tabular}
    \caption{$\mathcal{M}$ is the initial turbulent Mach number and $\eta_R$ the reference ambipolar diffusion coefficient.}
    \label{tab:para}
\end{table}

The ambipolar diffusivity $\eta_A$ in the governing equations is uncertain. It is strongly affected by the size distribution of dust grains and the abundance of metals (such as Mg) in the gas phase, which are uncertain \citep[e.g.][]{Lesur2021_review}. Following \citet{Lam2019}, we choose to parameterize it in the following way for efficient parameter exploration: 
\begin{equation}
    \eta_A = \frac{B^2}{4\pi\gamma\rho\rho_i}\label{equ:ambipolar_coeff},
\end{equation}
where $\gamma=\langle\sigma v\rangle/(m + m_i)$ is the ion-neutral drag coefficient and $\rho_i = C\rho^{1/2}$ is the ion density, with the coefficient $C$ proportional to the square root of the cosmic ray ionization rate \citep{Shu1992,Draine1983}; we postpone a more detailed treatment of the ambipolar diffusivity using chemical networks to a future investigation. With these approximations, the ambipolar diffusivity can be written as
\begin{equation}
    \eta_A = \eta_0 \frac{B^2}{4\pi\rho^{3/2}},
    \label{equ:etaq}
\end{equation}
with $\eta_0 = 1/(\gamma C)$. For the standard cosmic ionization rate of $10^{-17}$~s$^{-1}$, 
we have $\gamma=3.5\times 10^{13}\mathrm{cm^3g^{-1}s^{-1}}$ and $C=3\times10^{-16}~{\rm cm}^{-3/2}~{\rm g}^{1/2}$ \citep{Shu1992}, yielding a ``reference" value for $\eta_0$ of $\eta_R=95.2~\mathrm{g^{1/2}~{\rm cm}^{-3/2}~{\rm s}}$. Besides $\eta_R$, we also explore a wide range of values for $\eta_0$, from 0.1 to 10~$\times~\eta_R$ (see Table~\ref{tab:para}).  

We choose 6 levels of mesh refinement, with the finest cell size of 1.22~au at the highest level of refinement.  We ensure that the sink particle always stays on the finest grid so that the resolution for the protostellar disk, if formed, is maximized.

\begin{figure*}
    \centering
    \includegraphics[width=\textwidth]{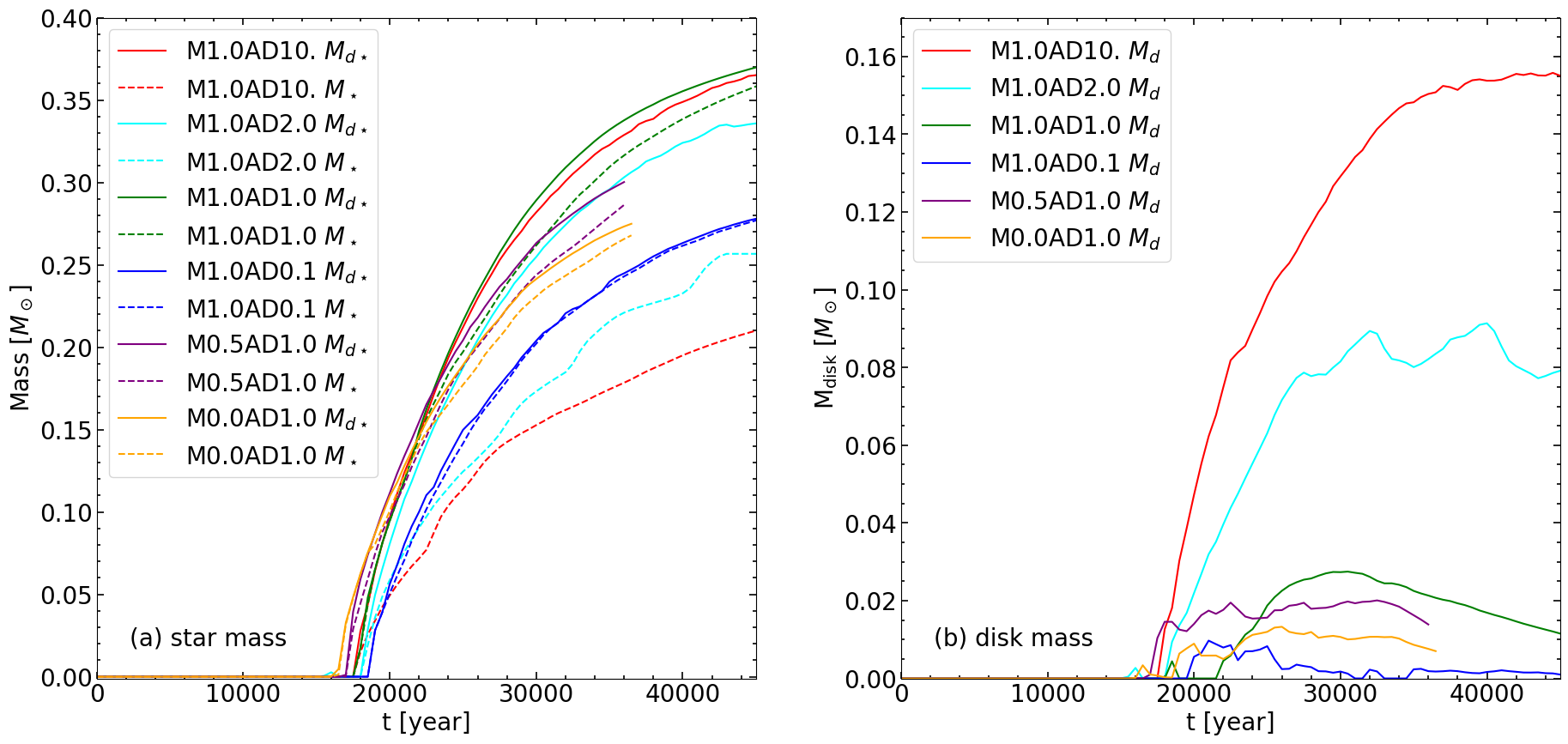}
    \caption{Time evolution of the stellar mass $M_*$ (panel a; dashed lines) and disk mass $M_d$ (panel b) as a function of time for all six models listed in Table~\ref{tab:para}.  Also plotted in panel (a) is the combined star and disk mass $M_{d*}$ (solid lines). 
    }
    \label{fig:overview}
\end{figure*}

\begin{figure*}
    \centering
    \includegraphics[width=\textwidth]{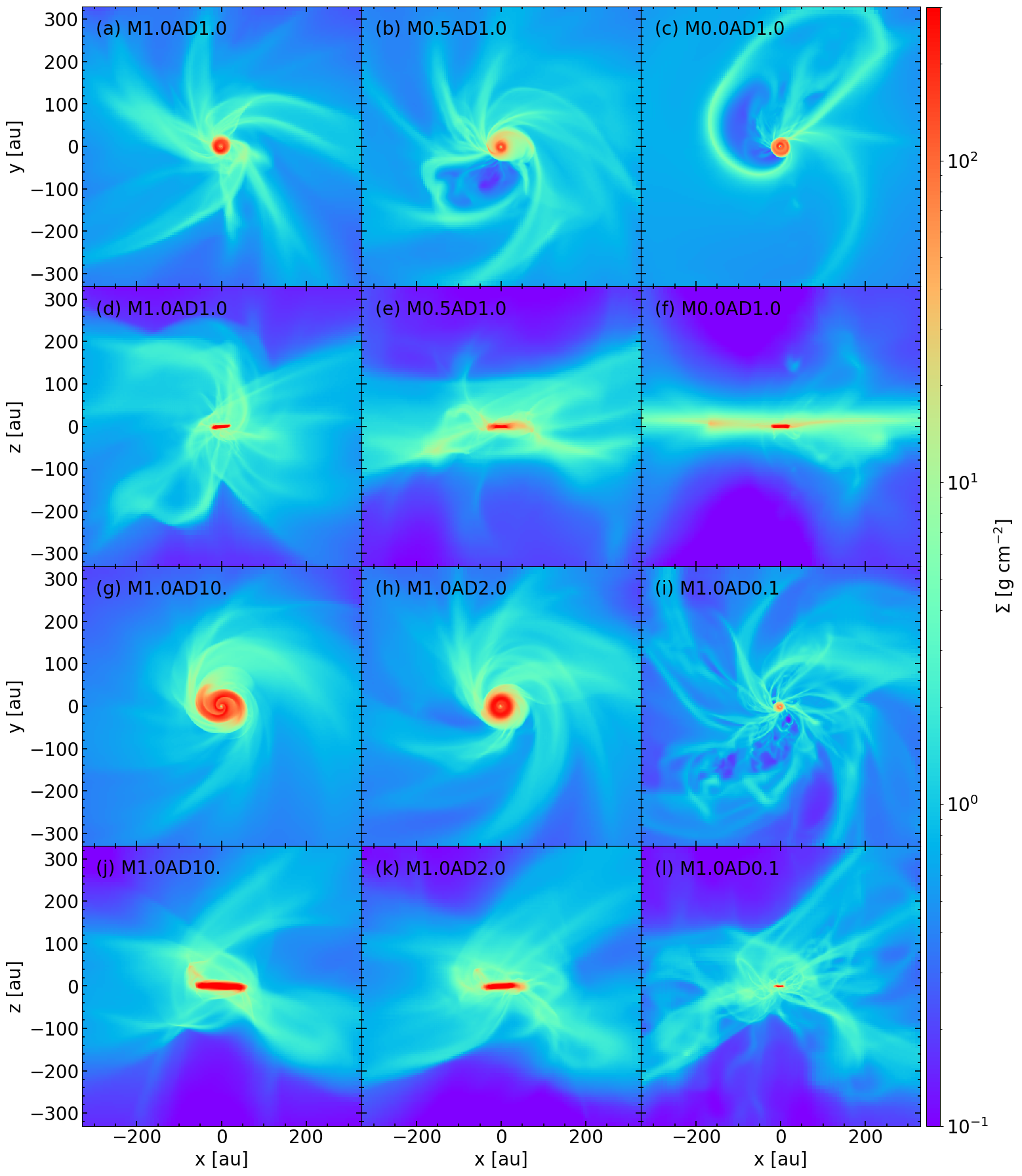}
    \caption{Face-on (first and third row) and edge-on (second and fourth row) views of the column density distributions of all six models listed in Table~\ref{tab:para}, showing the formation of a dense disk surrounded by a filamentary inner envelope at a representative time when the combined mass of the star and disk $M_{d*}=0.2$~M$_\odot$. The model name is shown in each panel. 
    The animated version of this figure is available at \url{https://figshare.com/s/4b8822b823a7ded2d9ba}.
    }
    \label{fig:col_den}
\end{figure*}

\section{Overview of Simulation Results}
\label{sec:simulation_result_overview}

To give a first impression of the simulation results, we show in Figs.~1 and 2 some broad features of all six models listed in Table~\ref{tab:para}. Fig.~1 plots the masses of the protostar (sink particle, panel a) and the protostellar disk (panel b) as a function of time, and Fig.~2 displays the face-on (along the $z-$axis) and edge-on views of the column density distribution, showing a dense, flatted disk-like structure surrounded by a filamentary inner envelope in all cases. The disk is persistent and well defined in some cases (e.g., Models M1.0AD2.0 and M1.0AD10) but highly transient and irregular in others (e.g., Model M1.0AD0.1; see the animated version of Fig.~\ref{fig:col_den} in the supplementary material). 

To facilitate a quantitative analysis, we define the disk in two steps. Firstly, we identify a disk ``midplane,'' 
which satisfies the following two criteria: 
(a) it passes through the protostar (the sink particle), and (b) its average distance to the densest 10,000 cells in the simulation is minimized. A region on this plane is considered to be part of the disk if it is rotationally supported with $v_\phi > 0.8~v_\mathrm{K}$ where $v_\mathrm{K}$ is the local Keplerian speed) and denser than $\rho_\mathrm{disk}$. We have empirically varied the threshold density $\rho_\mathrm{disk}$ and found that a value of $5\times 10^{-15}$ g/cm$^3$ describes well the boundaries of the relatively small disks formed in all cases other than the two most magnetically diffusive models (M1.0AD2.0 and M1.0AD10), where the disks are substantially larger and less dense (see Fig.~\ref{fig:col_den} below). For these two cases, we apply a lower density threshold of $1\times 10^{-15}$ g/cm$^3$, which captures the full extents of the larger and lower-density disks better (see, e.g., Fig.~\ref{fig:toomerQ}).
%
%
Secondly, we identify the rest of the disk by examining regions on planes parallel to, and at increasing distances from, the midplane. A region on a plane is considered part of the disk if (a) its density is greater than $\rho_\mathrm{disk}$, (b) its location lies within the disk boundary when projected onto the midplane, (c) its rotational motion dominates over the vertical motion, with $v_\phi > 2\ |v_z|$ (following \citealt{Masson2016}). In addition, we require that the disk cells form nearly closed circles around the midplane axis, to exclude irregular, rotationally dominated accretion streams that can extend tens of AUs above and below the disk midplane; these streams can be seen in, e.g., panel (d) of Fig.~\ref{fig:col_den} and will be discussed in \S~\ref{sec:Envelope2Disk} below. 

We can quantify the disk's mass with the above definition, as shown in Fig.~\ref{fig:overview}b. The disk mass is determined primarily by the magnetic diffusivity, with more massive disks formed in the more diffusive models (M1.0AD10 and M1.0AD2.0). The more massive disks are also larger, as shown in panels (g) and (h) of Fig.~\ref{fig:col_den}. Conversely, the less diffusive (i.e., best magnetically coupled) model, M1.0AD0.1, barely has a rotationally supported disk (see Fig.~\ref{fig:col_den}i), especially at late times, when the disk mass goes to almost zero (see the blue curve in Fig.~\ref{fig:overview}b). The reference run, M1.0AD1.0, lies between these two extremes, with a well-defined disk over most of the simulation duration, although both its mass and radius decrease at later times. As we will show in \S~\ref{sec:DiskOnly} below, the disk formed in this case evolves quickly, with the material accreted from the envelope passing through the disk in only several orbital periods.

Compared to the disk mass $M_d$, the stellar mass $M_*$ varies less across the different models, as shown by the dashed lines in Fig.~\ref{fig:overview}a. The difference between the largest $M_*$ (for the reference model M1.0AD1.0) and the least massive one (for the most diffusive model M1.0AD10) is less than a factor of two (compare the green and red dashed curves in the panel). The difference is even less for the combined stellar and disk mass $M_{d*}$ because smaller stellar masses in more diffusive models are compensated somewhat by larger disk masses (see the solid lines in the panel). Nevertheless, there is some difference in $M_{d*}$ across the models, with the largest value in the reference model and the lowest value in the best magnetically coupled model M1.0AD0.1, indicating a difference in the structure and dynamics of the protostellar envelope, which controls the envelope accretion onto the disk-plus-star system, and/or in the outflow rate.

We discuss the structure and dynamics of the protostellar envelope in detail in the next section (\S~\ref{sec:envelope}). It is followed by a discussion of the connection between the envelope and the disk in \S~\ref{sec:Envelope2Disk} and the structure and evolution of the formed disk in \S~\ref{sec:DiskOnly}.

\section{Magnetically Structured Protostellar Envelopes}
\label{sec:envelope}

\begin{figure*}
    \centering
    \begin{minipage}{0.75\textwidth}
        \includegraphics[width=\textwidth]{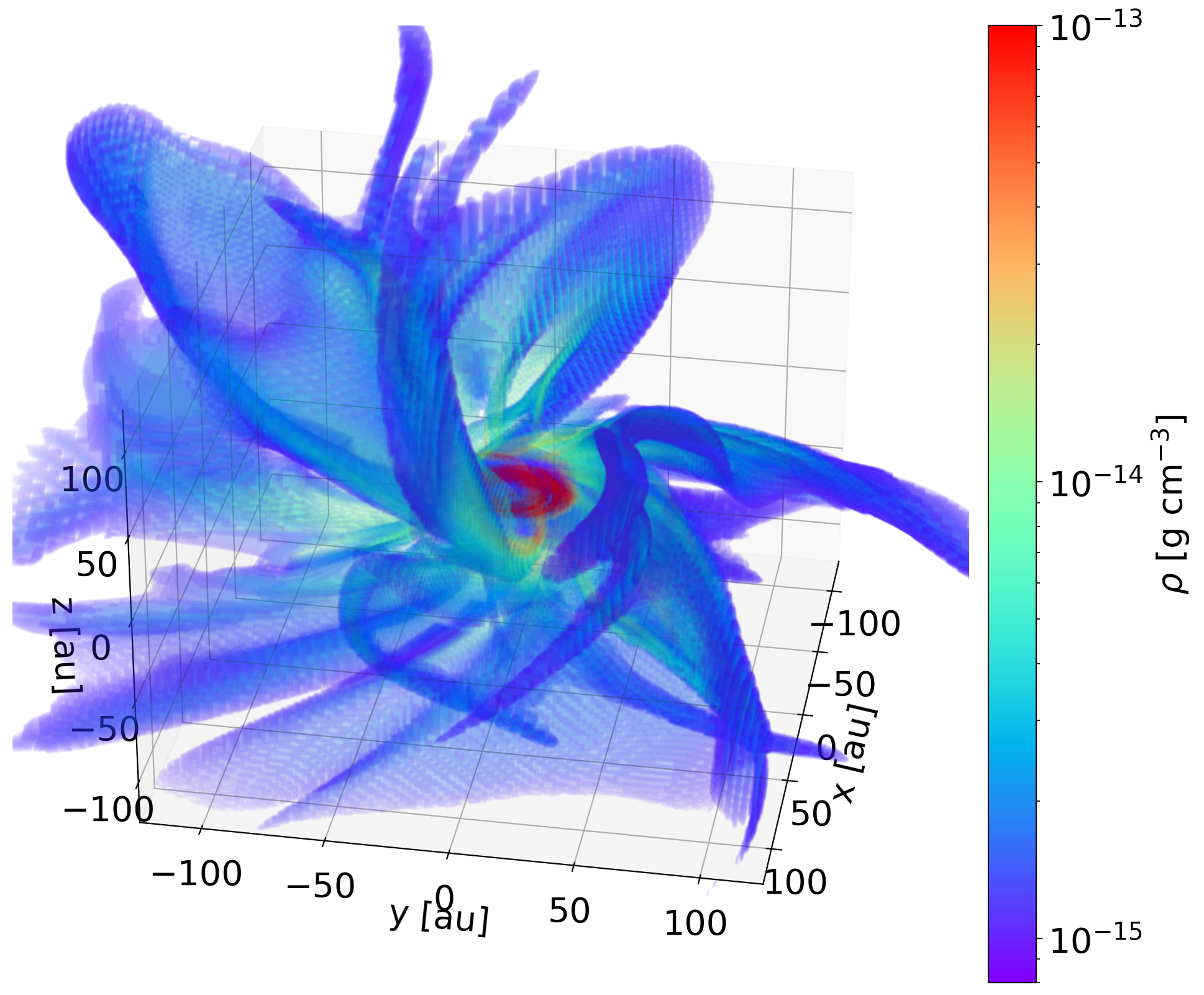}
    \end{minipage}
    \caption{3D visualization of the reference model (M1.0AD1.0) showing that most of the inner envelope mass is concentrated into warped but coherent sheet-like structures (termed ``gravo-magneto-sheetlets" or simply ``sheetlets" in the text). 
    The animated version is available at  \url{https://figshare.com/s/1bf4ed63e345d01c157a}.
    }
    \label{fig:3D}
\end{figure*}

As mentioned in the introduction, one of our focuses is on how the interplay between the magnetic field, turbulence, and gravity shapes the structure and dynamics of the protostellar envelope. 
The inner protostellar envelopes are highly structured, as illustrated by the representative column density maps shown in Fig.~\ref{fig:col_den}. To highlight the envelope, we pick a representative time that is relatively early when the combined stellar and disk mass reaches $M_{d*}=0.2$~$M_\odot$ in this section (\S~\ref{sec:envelope}). 
 The envelope becomes more inhomogeneous and filamentary with an increasing initial turbulence strength (compare Models M1.0AD1.0, M0.5AD1.0, and M0.0AD1.0). In the non-turbulent model M0.0AD1.0 (panels c and f), the gas in the envelope concentrates near the midplane, forming an infalling disk-like structure that is not rotationally supported -- the pseudodisk\footnote{The low-density cavity in Fig.~\ref{fig:col_den}c is filled with the magnetic flux released by the mass already accreted to the central protostar, as first discussed in \citet{Zhao2011}.
 %
 %
 }. In the more turbulent model M0.5AD1.0, the pseudodisk puffs up into a visually filamentary structure with a vertical height of about 150~au at the time shown (panel e). In the most turbulent reference model M1.0AD1.0, the envelope material falls onto the disk from all directions in dense filaments (panel d).

Although the dense filaments in the column density maps of the reference model appear to be independent of one another, they are spatially connected in 3D.  The spatial connectivity is illustrated in  Fig.~\ref{fig:3D}, where we highlight the high-density regions in 3D with colored dots at the same time as in Fig.~\ref{fig:col_den}. The figure shows only the densest regions above a density threshold $\rho_{c} (r)$ that contains 70\% of the mass at each radius $r$ (see more discussion in \S~\ref{subsec:role} below). Regions with densities below $4\times10^{-16}\ \mathrm{g\ cm^{-3}}$ are removed from the visualization for clarity. Most of the inner envelope material is concentrated into warped sheet-like structures.
These 3D structures can be seen even more clearly in the animation included in the online supplementary materials. They show up as the long, curved filaments in the (projected) column density maps (see Fig.~\ref{fig:col_den}a,d)\footnote{We should note that some filamentary structures are produced even without an initial turbulence. For example, there is an elongated dense loop in the non-turbulent model M0.0AD1.0 that lines the outer edge of the low-density cavity carved out by the escaping magnetic flux in the equatorial pseudodisk \citep[see Fig.~\ref{fig:col_den}c and also the lower-left panel of Fig.~2 of][]{Mignon-Risse2021}. It appears different from the filamentary structures produced by the initial turbulence, which warps (rather than carving out) the pseudodisk. The turbulence-induced sheet-like structures tend to be farther away from the equatorial plane (see Fig.~\ref{fig:col_den}d and more pictorially Fig.~\ref{fig:3D}) and generally do not line the edges of any obvious cavities driven by the released magnetic flux. It is conceivable that the released flux could still have an effect on the inner envelope in the presence of an initial turbulence, but it is unlikely to dominate the creation of the dense sheet-like structures shown in Fig.~\ref{fig:3D}, which look very different from the equatorial dense structure created by the released flux in the non-turbulent case. 
}. 

The dense warped sheet-like structures also exist in models with different ambipolar diffusivities, as shown in the column density maps of Models M1.0AD10, M1.0AD2.0, and M1.0AD0.1 in Fig.~\ref{fig:col_den}.
There is a clear trend that the inner envelope is more inhomogeneous and filamentary when the magnetic field is better coupled to the gas (i.e., for a lower ambipolar diffusivity). Since the ambipolar diffusivity controls the efficiency of magnetic braking and the disk size (as discussed in detail in Section~\ref{sec:Envelope2Disk} below), there is a correlation between the disk size and how filamentary the envelope is, everything else being equal; such a correlation should be searched for in observational data.
The sheet-like structures appear to be a generic feature of the turbulent magnetized inner protostellar envelopes. Their ubiquitous presence makes it conceptually appealing to view the inner envelope as a bi-phase medium, with dense sheet-like structures embedded in a low-density background. 

\subsection{Dense Collapsing Gravo-Magneto-Sheetlets}
\label{subsec:sheetlet}

\begin{figure*}
    \centering
    \includegraphics[width=\textwidth]{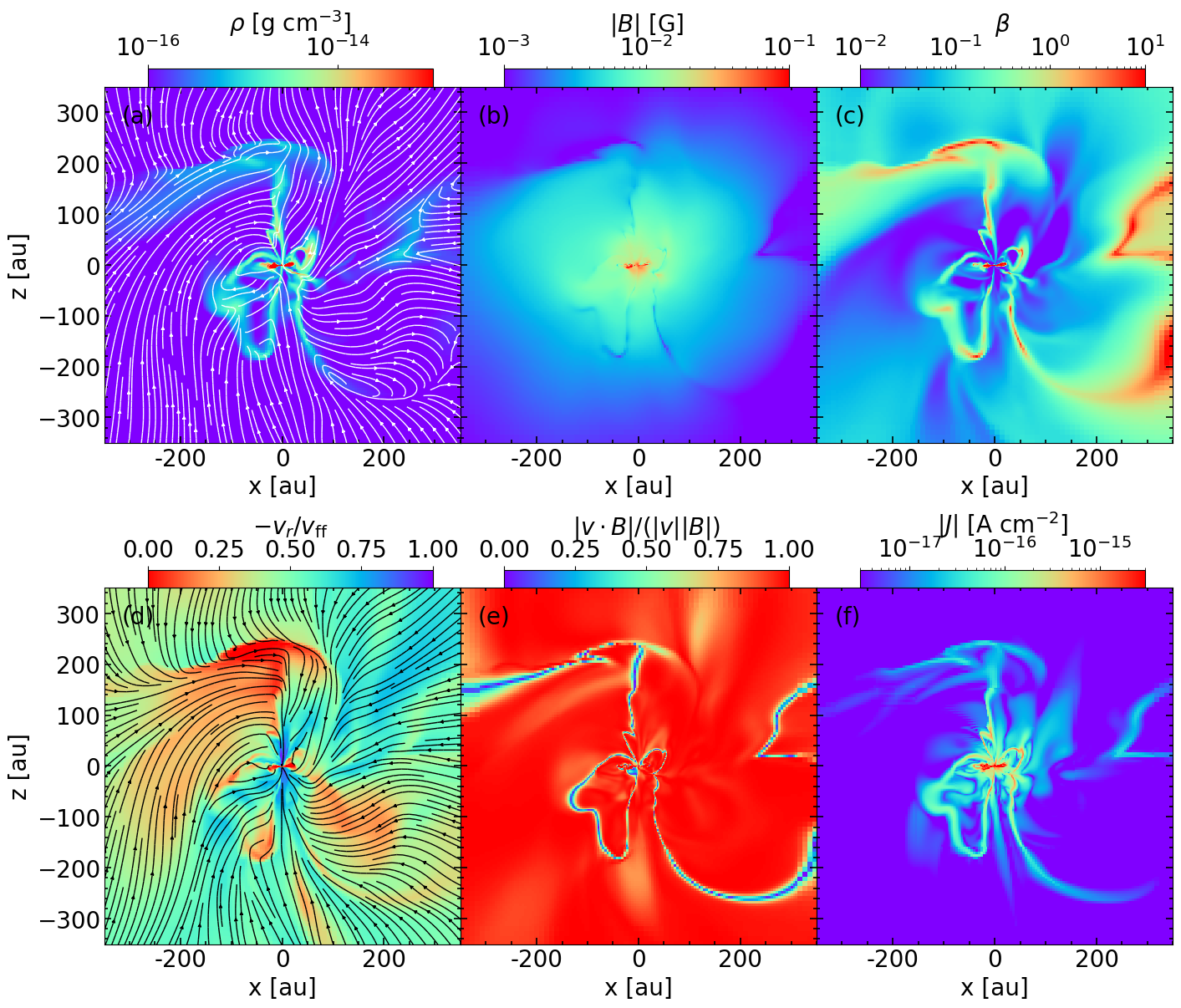}
    \caption{Illustration of a highly inhomogeneous magnetically-structured protostellar envelope. Panel (a) plots the mass density $\rho$ on an $x$-$z$ plane, showing thin dense filaments (sheetlets in 3D) embedded in a low-density background. The field lines (white) demonstrate that the dense structures are associated with strong field pinching, as in the case of an unperturbed pseudo-disk. Panel (b) shows that the density increase in the dense structures does not have a corresponding increase in the total field strength. Panel (c) displays the plasma-$\beta$, showing the dense structures are much less magnetized relative to their masses than the background. Panel (d) shows that the envelope tends to collapse radially at a sub-free-fall speed (with flow streamlines superposed to illustrate the velocity field). Panel (e) displays the cosine of the angle between the velocity and magnetic field vectors, showing the magnetically guided flows in the low-density background and the gravity-driven collapse in the cross-field direction in the dense structures. Panel (f) displays the distribution of the current density $\vert\mathbf{J}\vert$, showing that the current (and thus the magnetic force) is concentrated in the dense structures. 
    The animated version is available at  \url{https://figshare.com/s/a4e0d11512da5dbf911b}.
    }
    \label{fig:slices}
\end{figure*}

To help illustrate the physical nature of the dense sheet-like structures and their differences from the lower-density background, we plot in Fig.~\ref{fig:slices}a the distribution of the mass density $\rho$ on an $x$-$z$ plane passing through the stellar position for the reference model M1.0AD1.0 at the representative time. It can be seen that dense filaments clearly stand out from the lower-density background. They are the cross-sections of the 3D dense sheet-like structures shown in Fig.~\ref{fig:3D} on the chosen plane. Several magnetic field line segments are superposed on the density map to demonstrate that the dense structures are associated with a strong field line pinching, which is a well-known characteristic of an unperturbed magnetically induced pseudodisk (\citealp{Galli&Shu1993}; see, e.g., Fig.~6 of \citealp{Li2014}). The similarity in the local field geometry supports the notion that the dense sheet-like structures are distorted pseudodisks, which are produced by the gravitational accumulation of material along the field lines near the sharp kinks where the gravitational collapse is retarded by the strong magnetic tension force associated with the field pinching. Just as in the case of unperturbed equatorial pseudodisk, the formation of these dense structures is an unavoidable consequence of the intrinsically anisotropic magnetic resistance (mostly through tension forces) to the gravitational collapse. To highlight their common physical origin with the unperturbed pseudodisks that are sheets produced by the interplay between gravitational collapse and anisotropic magnetic resistance, we will term these dense structures ``gravo-magneto-sheetlets" or simply ``sheetlets" for short hereafter. They are simply pseudodisks distorted and/or disrupted by turbulence or other processes (such as a clumpy mass distribution). 

Because the dense sheetlets/filaments are formed by mass accumulation primarily along the field lines, 
the large density increase is not accompanied by a correspondingly large increase in the field strength, as illustrated in panel (b) of Fig.~\ref{fig:slices}. If anything, the field strength in the dense filaments tends to be lower than their surroundings (see, e.g., the two filaments below the disk). It is a direct consequence of the sharp field line pinch across the dense filament, which greatly increases the component of the magnetic field parallel to the filament immediately outside it compared to that inside. The combination of higher densities and comparable or weaker magnetic fields makes the dense structures easily distinguishable from their background in the plasma-$\beta$, as shown in panel (c). It is clear that, whereas the background is strongly magnetically dominated, with a plasma-$\beta$ well below unity, the dense structures are much less so, with a plasma-$\beta$ closer to, or, in some regions, even larger than unity. This difference re-enforces the notion of a bi-phase inner envelope where a magnetically dominated background surrounds a less magnetized (relative to the local density) network of denser filaments (or sheetlets in 3D).

The strong magnetic field in the inner envelope, especially in the low-density background, is expected to affect gas kinematics. Naively, one would expect the inner envelope to collapse more or less radially towards the central stellar object at a speed close to the local free-fall value, with some deflection towards the midplane near the disk because of rotation. However, this is not the case, as illustrated in panel (d), where the streamlines with velocity vectors (black arrows) are displayed. There is a large deviation from the expected radial infall over most of the displayed area, with non-radial motions dominating the radial infall over large patches. It is particularly true in the low plasma-$\beta$ regions to the upper-left and lower-right of the disk, where the flows are more azimuthal than radial. They broadly follow the magnetic field lines (shown in panel a), indicating that the envelope material in these regions is forced to slide along the rather strong (and thus rigid) field lines, which are not directed radially.

Another major deviation from the naive expectation is that the radial collapse speed is typically significantly below the local free-fall speed (defined as $\sqrt{2\vert \Phi\vert}$ where $\Phi$ is the local gravitation potential) in the inner envelope (see the color map in panel d), by a factor of 3 or more over large patches. Since these slowly infalling regions do not have a significant amount of rotational support, their gravitational collapse must be retarded primarily by magnetic forces. The magnetic retardation will be discussed further in Section~\ref{sec:mag_forces}, and its observation implications will be discussed in Section~\ref{sec:discussion}. 

The most striking difference between the dense sheetlets and the background is the degree of alignment between the flow and magnetic field directions (see panel e). Whereas the low-density gas in the magnetically-dominated (low-plasma-$\beta$) background primarily flows along the field lines (with $\cos[\theta]\equiv \vert \mathbf{v}\cdot\mathbf{B}\vert/[\vert \mathbf{v}\vert~\vert\mathbf{B}\vert]$ close to unity, where $\theta$ is the angle between the velocity and magnetic vectors), the dense sheetlets move primarily perpendicular to the field lines. The bimodal behavior leads to a conceptual picture where the envelope material in the low-density background is collected along the (strong) field lines into the dense sheetlets, where the gravity becomes strong enough to overwhelm the magnetic (tension) resistance and pull the gas towards the central protostar in a cross-field direction, pinching the field in the process. In this picture, the gravity-dominated dense sheetlets are the main conduits for the envelope to feed the central disk-plus-star system, while the magnetically-dominated background is the feeder of material to the conduits, analogous to the runoff of rainwater from the land surface feeding rivers. In other words, the magnetized protostellar envelope accretion is fundamentally a two-step process involving two distinct types of regions shaped by the competition between gravity and magnetism -- dense gravity-dominated sheetlets and diffuse magnetically-dominated background -- with the latter feeding the former, which in turn feeds the central star-disk system. We call this two-step feeding process ``bimodal protostellar envelope accretion."
%
%

The dense sheetlets/filaments and the background also differ in the current density $\mathbf{J}$, which is displayed in panel (f) of Fig.~\ref{fig:slices}. The current density map looks strikingly similar to the mass density map (panel a), with large current densities concentrating in the dense filaments. The strong correlation is not surprising since the dense filaments are where the field lines are sharply pinched (see panel a). An implication is that the magnetic force ($\propto \mathbf{J}\times\mathbf{B}$) is also concentrated in the dense filaments, including the azimuthal component, which is responsible for regulating the angular momentum content of the envelope through magnetic braking. We will return to discussing magnetic braking and disk formation in \S~\ref{sec:Envelope2Disk}. 

It should be noted that the dense gravo-magneto-sheetlets can be modified by the magnetic flux released near the center, which tends to expand against (and sweep up) the surrounding material, including the sheetlets. In the non-turbulent case (M0.0AD1.0), the magnetic flux release creates a ``bubble'' on the pseudodisk (fig.~\ref{fig:col_den}c), which aligns with the disk midplane. In the turbulent models, the magnetic flux release would interact with the dense sheetlets in a similar way, although the flux release is more episodic and harder to quantify.

\begin{figure*}
    \centering
    \includegraphics[width=\textwidth]{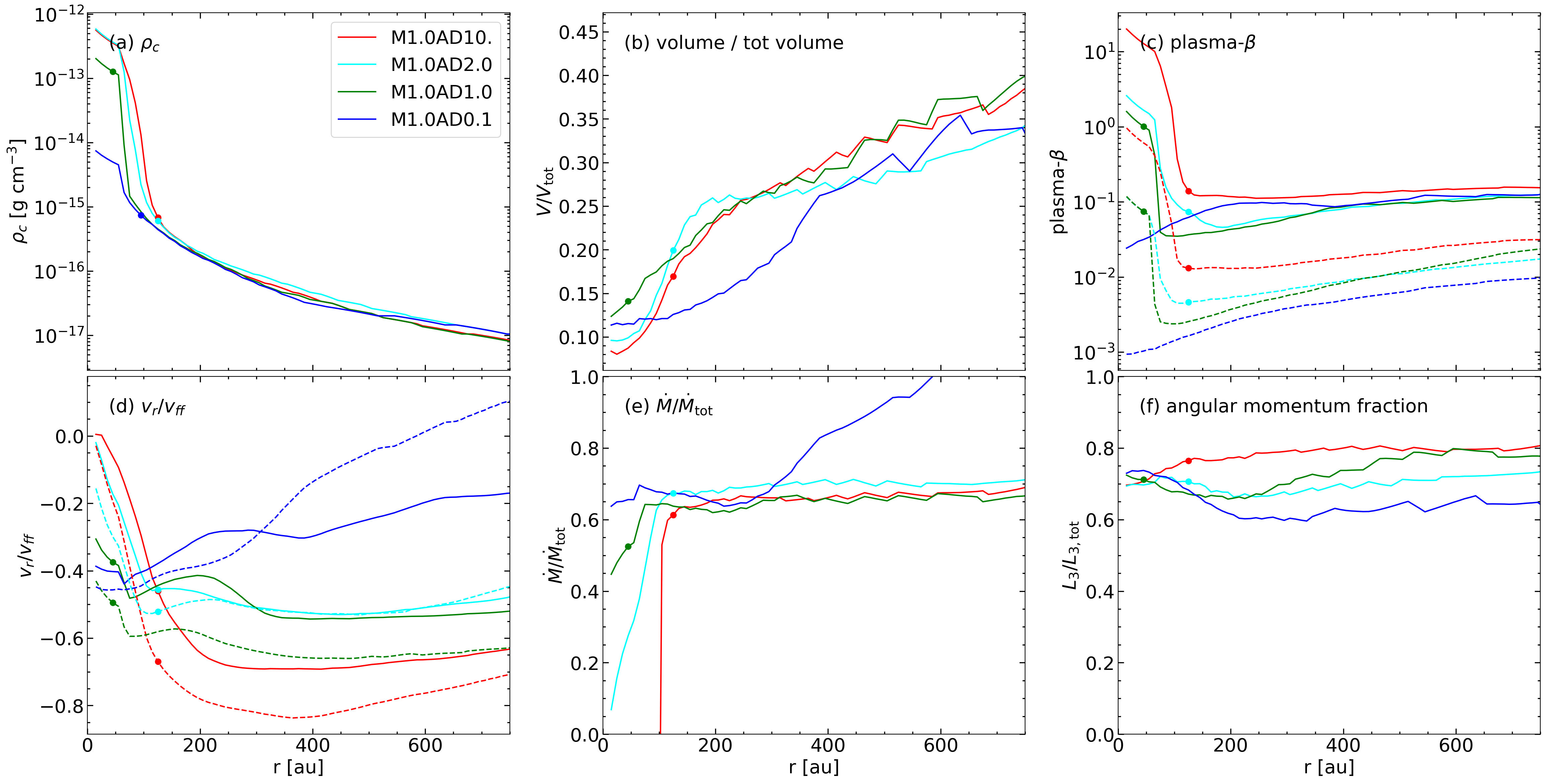}
    \caption{Spherically average properties of dense sheetlets and their contributions to the structure and dynamics of the inner protostellar envelope for models with different ambipolar diffusivities. Panel (a) displays the characteristic density $\rho_{c}$ that separates the dense sheetlets from the diffuse background. Panel (b) shows that the dense sheetlets occupy an increasingly small fraction of the volume towards the center, reaching values as low as $\sim 10\%$. Panel (c) displays the plasma-$\beta$ for the sheetlets (solid lines) and background (dashed), showing an order-of-magnitude difference between the two. 
    Panel (d) plots the radial infall speed normalized by the local free-fall speed,
    showing that both the sheetlets and background collapse radially at substantially sub-free-fall speeds. Panel (e) shows the fraction of the radial mass flux carried by the dense sheetlets. Panel (f) displays the fraction of the $z-$component of the angular momentum carried by the dense sheetlets. The solid dot on each line approximately marks the transition radius between the disk and envelope on the disk midplane.}
    \label{fig:back_bones}
\end{figure*}

\subsection{Role of Sheetlets in Envelope Structure and Dynamics}
\label{subsec:role}

To quantify the roles of the dense (3D) sheetlets in the structure and dynamics of the protostellar envelope, we define a characteristic density $\rho_{c} (r)$ at each radius $r$ from the central protostar, with the material denser than $\rho_{c}$ containing $70\%$ of the mass at that radius. As shown in Fig.~\ref{fig:3D}, the material above this density threshold forms multiple sheet-like structures in the reference model -- the dense sheetlets. The characteristic density is shown as a function of radius in the inner envelope up to $750$~au in Fig.~\ref{fig:back_bones}a for models with the same initially sonic turbulence but different ambipolar diffusivities at the representative time when $M_{d*}=0.2$~M$_\odot$. The sharp increase in $\rho_{c}$ towards the center on each curve marks the transition from the envelope to the disk. 
The dense sheetlets defined this way contain the majority of the envelope mass. However, they are squeezed into an increasingly small fraction of the volume towards the center, reaching values as low as $\sim 10\%$, as shown in panel (b). 
Panel (c) displays the average plasma-$\beta$ for the dense sheetlets and diffuse background, showing an order-of-magnitude difference between the two, especially for the least magnetically diffusive model M1.0AD0.1, where the contrast is the highest. 
Panel (d) plots the radial infall speed normalized by the local free-fall speed. For the reference model M1.0AD1.0, the panel shows that the dense sheetlets collapse at roughly $50\%$ of the local free-fall speed, somewhat slower than the diffuse background, which infalls at about $60\%$ of the free-fall speed. The sub-free-fall radial motions are consistent with the normalized infall speed shown pictorially in the color map of  Fig.~\ref{fig:slices}d. The infall speed tends to decrease with a smaller magnetic diffusivity. It remains substantially sub-free-fall for all four models shown, indicating that the result is robust. 
It is primarily a consequence of the magnetic field channeling the flow in the low-plasma $\beta$ background and resisting gravitational collapse in dense sheetlets, which will be discussed in more detail in \S~\ref{sec:Envelope2Disk}.

Because the average infall speeds are comparable in the dense sheetlets and diffuse background, the former (which contains 70\% of the mass by definition) typically contributes the majority ($\sim 60-70\%$) of the (radial) mass accretion rate in the M1.0AD1.0, M1.0AD2.0, and M1.0AD10 models. The mass accretion contribution by the sheetlets in the lowest diffusivity model M1.0AD0.1 is larger beyond a radius of $\sim 300$~au, reaching values greater than 100\% because the background has a positive average radial speed at such large radii (see panel d) and thus contributes negatively to the mass accretion. 
Similarly, the dense sheetlets carry the majority ($\sim 60\%$ or more) of the $z-$component of the angular momentum in the envelope in all four cases, as shown in panel (f) of Fig.~\ref{fig:back_bones}, and are thus expected to play a crucial role in disk formation, as discussed further below in \S~\ref{sec:Envelope2Disk}. 

We conclude that, despite the small volume they occupy, the dense sheetlets play a key role in the structure and dynamics of the protostellar envelope, dominating its mass and angular momentum budgets and mass accretion rate.

\begin{figure*}
    \centering
    \includegraphics[width=\textwidth]{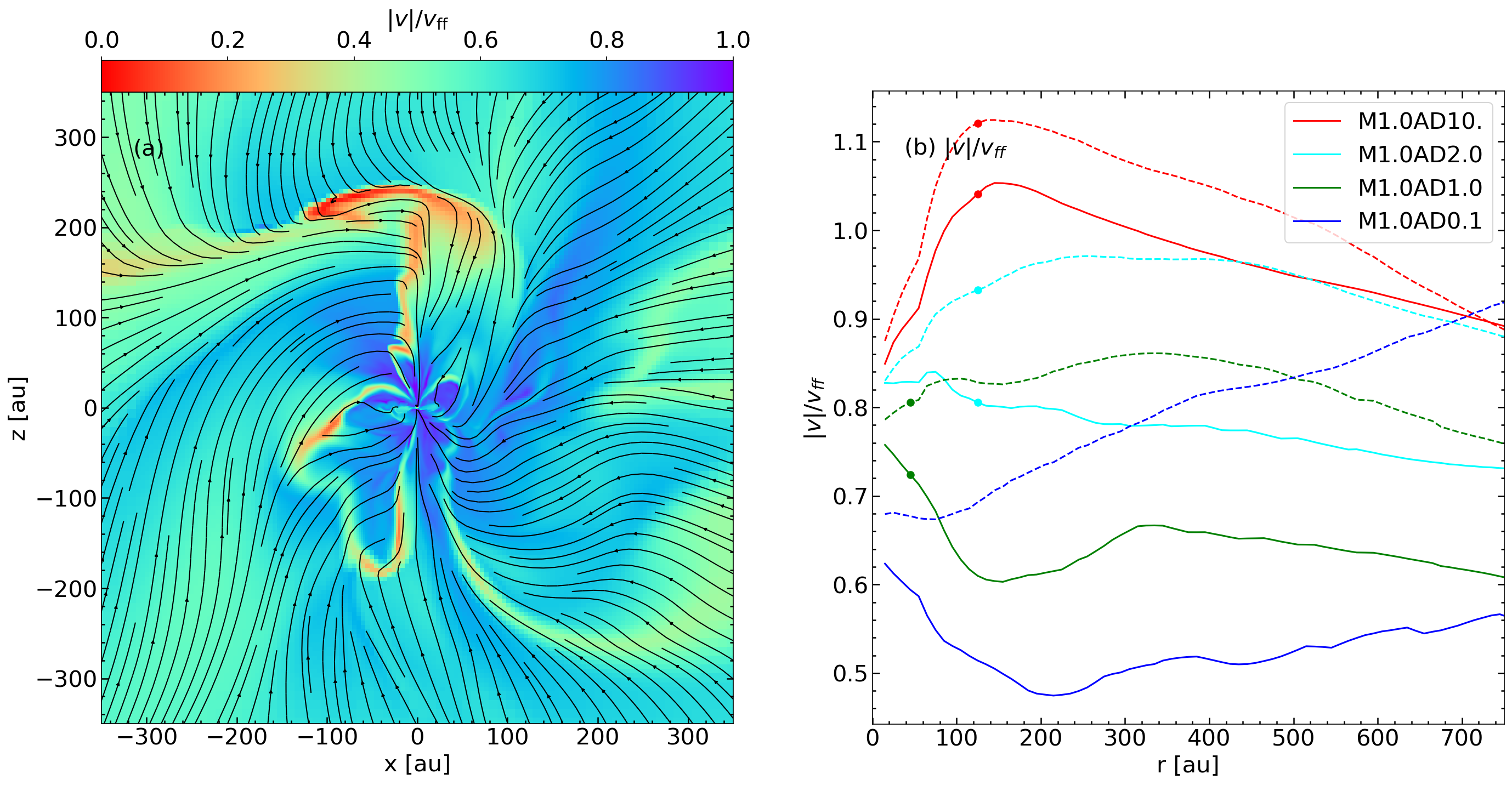}
    \caption{Comparison of the total flow speed $v$ with the local free-fall speed $v_{\rm ff}$. Panel (a) plots the ratio $v/v_{\rm ff}$ on an $x$-$z$ plane for the reference model M1.0AD1.0 at the same representative time as in Fig.~\ref{fig:slices}, showing dense sheetlets tend to move more slowly than the background medium. Panel (b) plots the spherically averaged total speed normalized by the local free-fall speed as a function of radius for models with different diffusivities. The blue lines are for model M1.0AD0.1, green for M1.0AD1.0, cyan for M1.0AD2.0, and red for M1.0AD10. The solid curves are for the dense sheetlets and dashed for the background. The solid dot on each line approximately marks the transition radius between the disk and envelope on the disk midplane.} 
    \label{fig:TotalSpeed}
\end{figure*}

Before leaving this subsection, we comment on a few interesting features regarding the total flow speed $v$ compared to the local free-fall speed $v_{\rm ff}$. The ratio $v/v_{\rm ff}$ is plotted on an $x$-$z$ plane for the reference model M1.0AD1.0 at the same representative time as in Fig.~\ref{fig:TotalSpeed}, so it can be compared directly with Fig.~\ref{fig:slices}d, where the radial component of the velocity normalized by the local free-fall speed is plotted. The contrast between the dense sheetlets and the more diffuse background is more clear in the total speed plot than in the radial velocity plot. Specifically, the sheetlets tend to move significantly slower than the background (with a redder color in Fig.~\ref{fig:TotalSpeed}a) and at a substantially sub-free-fall speed. The difference is quantified in panel (b) of Fig.~\ref{fig:TotalSpeed}, where the spherically averaged total speed normalized by the local free-fall speed is plotted as a function of radius for models with different diffusivities. There is a clear trend for the dense sheetlets to move more slowly for a lower magnetic diffusivity (compare the four solid lines in the panel). In the least magnetically diffusive (best coupled) case of M1.0AD0.1, the sheetlets move at about half of the local free-fall speed on average. They move close to the free-fall speed in the most diffusive case (Model M1.0AD10). Another interesting trend is that the difference in the total speed between the dense sheetlets and their lower-density background is larger for a lower magnetic diffusivity (contrast, e.g., the blue curves with the red ones), which, together with a smaller volume filling factor (see Fig.~\ref{fig:back_bones}b) and higher contrast in plasma-$\beta$ (see Fig.~\ref{fig:back_bones}c), indicate that the sheetlets become more distinct from their lower-density background as the magnetic field becomes better coupled to the gas. 

\section{The Envelope-Disk Connection}
\label{sec:Envelope2Disk}

In the last section, we demonstrated the importance of the magnetic field in segregating the envelope material into dense gravo-magneto-sheetlets and a magnetically dominated low-density background. In this  section, we first examine the field strength and distribution in some detail, focusing on the vertical field component $B_z$  (\S~\ref{sec:level_of_magnetization}), which controls the {\it net} magnetic flux threading the formed disk $\Psi_z$ (a crucial quantity for the subsequent disk dynamics and evolution), and the connection between the inner envelope and the disk. It is followed by a discussion of magnetic retardation of the envelope infall and magnetic braking of rotation in \S~\ref{sec:mag_forces}, and anisotropic (sheetlet-fed) envelope accretion onto the disk in \S~\ref{sec:accretion}. In this (\S~\ref{sec:Envelope2Disk}) and most of the next sections  (\S~\ref{sec:disk_gravity_stability} and \S~\ref{sec:disk_ang_mom_transport}), we choose to illustrate the results at a representative time when the stellar and disk mass reaches $M_{d*}=0.25~M_\odot$ (somewhat later than that used in the previous envelope-focused section corresponding to $M_{d*}=0.2~M_\odot$) to have a larger, more stable disk that facilitates the disk discussion. 

\subsection{Level of Envelope and Disk Magnetization}
\label{sec:level_of_magnetization}

\begin{figure*}
    \centering
    \includegraphics[width=\textwidth]{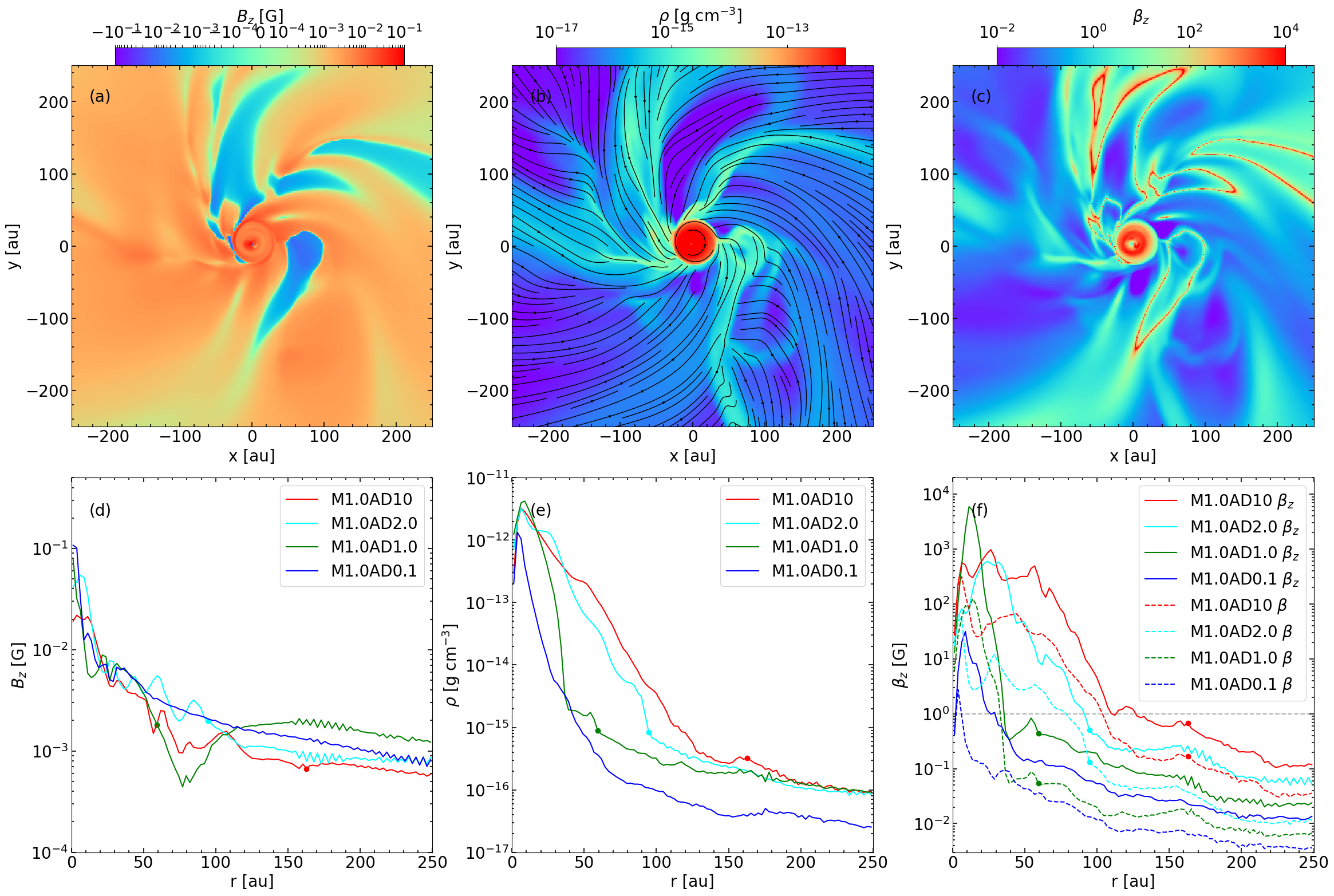}
    \caption{Magnetization of the inner envelope and disk at a representative time when $M_{d*}=0.25$~M$_\odot$ for the reference model M1.0AD1.0 (top panels) and for models with different magnetic diffusivities (bottom panels). Panel (a) displays the vertical component of the magnetic field on the midplane $B_{z,mid}$ for the reference model, showing the lack of a large jump in the disk-envelope transition region. Panel (b) is a density map on the same plane (with flow streamlines superposed), showing a dense disk surrounded by a much lower-density envelope. Panel (c) plots the plasma-$\beta_z$ corresponding to $B_{z,mid}$. Panels (d)-(f) plot, respectively, the azimuthally averaged $B_{z,mid}$, $\rho_{mid}$, and $\beta_{z,mid}$ for models of different magnetic diffusivities. The solid dot on each line approximately marks the transition radius between the disk and envelope. 
    The animated version is available at  \url{https://figshare.com/s/87c29218a42e2b711657}.
    }
    \label{fig:Bz_map}
\end{figure*}

To illustrate the magnetic field distribution, we plot in Fig.~\ref{fig:Bz_map}a the vertical field strength $B_{z,mid}$ on the $x$-$y$ plane passing through the sink particle (approximately the disk midplane) for the reference run M1.0AD1.0 at the representative time\footnote{The approximate alignment of the disk axis along the $z-$direction is likely due to the net angular momentum along the $z-$axis. The choice of the $x$-$y$ plane simplifies the computation of magnetic flux through a surface and best captures the flux of the original z-direction magnetic field.}.  Clearly, $B_{z,mid}$ varies strongly from one location to another in the inner envelope, which is to be expected given its highly structured mass distribution as discussed in \S~\ref{sec:envelope}. In particular, magnetic field lines are strongly distorted by the collapsing dense sheetlets, which can reverse the direction of the vertical B field in some locations (see, e.g., the field lines to the right of the right filament below the disk in Fig.~\ref{fig:slices}a). Such negative $B_z$ is also present on the midplane as shown in Fig.~\ref{fig:Bz_map}a. It would reduce the net vertical magnetic flux on the disk if advected to the disk along the midplane. 

Despite the strong spatial variation, there is a general trend for $B_{z,mid}$ to increase with decreasing radius towards the center (to be quantified below). Nevertheless, the increase is relatively gentle. In particular, there is no large jump from the infalling envelope to the rotationally supported disk, unlike the mass density on the same plane ($\rho_{mid}$, which is plotted in Fig.~\ref{fig:Bz_map}b for comparison. The dense curving filaments in the envelope in the density map are simply the midplane cross-sections of the 3D gravo-magneto-sheetlets (see Fig.~\ref{fig:3D}) that collapse towards the disk (see the flow streamlines). 
The difference between the distributions of $B_{z,mid}$ and $\rho_{mid}$ highlights that the large concentration of mass in the disk does not lead to the same degree of magnetic flux concentration there. It is further 
quantified in panel (c) of Fig.~\ref{fig:Bz_map}, where we plot the plasma-$\beta_{z,mid}$ based on the vertical component of the magnetic field $B_{z,mid}$ only. Although there is a strong spatial variation, especially in the inner envelope, $\beta_{z,mid}$ is, on average, much higher in the disk than in the envelope, reaching values as high as $10^4$. The large increase comes primarily from the jump in density rather than the field strength, indicating that the disk material is strongly demagnetized relative to its mass compared to the envelope feeding it. This demagnetization is the key for the disk to form in the first place and survive to later times, as discussed in, e.g., \citet{Hennebelle_2016} and recently reviewed by \citet{Tsukamoto2023}.
 
To quantify the magnetic field and mass distribution further, we azimuthally average $B_{z,mid}$ and $\rho_{mid}$ and show the results in the lower panels of Fig.~\ref{fig:Bz_map}. As expected, there is a clear change of the averaged midplane density around the disk-envelope transition region (marked by a solid dot), inside which the density starts to increase steeply with decreasing radius (except for the least magnetic diffusive model M1.0AD0.1 where a well-defined disk does not exist at the time shown and, thus, no solid dot marking disk-envelope transition is plotted). The behavior of the average midplane $B_{z,mid}$ is more complex, partly because negative values of $B_z$ exist in some regions (see panel a), which cause strong variations. The magnetic fields with alternating polarities lead, in particular, to the dip near 70~au for the reference model M1.0AD1.0 (the green curve in panel d). Nevertheless, the azimuthally averaged plasma-$\beta$ remains well behaved, with the thermally dominated disk region well separated from the magnetically dominated envelope region (panel f of Fig.~\ref{fig:Bz_map}). For the most magnetically diffusive model (M1.0AD10) with the best-resolved disk, there is a gradual increase of the averaged plasma-$\beta_{z, mid}$ towards the center, reaching a plateau of $\sim 400$. The second most diffusive model (M1.0AD2.0) reaches a similar peak value, although the peak is higher in the reference case. The disk plasma-$\beta$ based on the total magnetic pressure is lower, typically by about an order of magnitude, indicating that the toroidal magnetic field is stronger than the vertical field by a factor of several on average near the disk midplane. 
%
%

\begin{figure*}
    \centering
    \includegraphics[width=\textwidth]{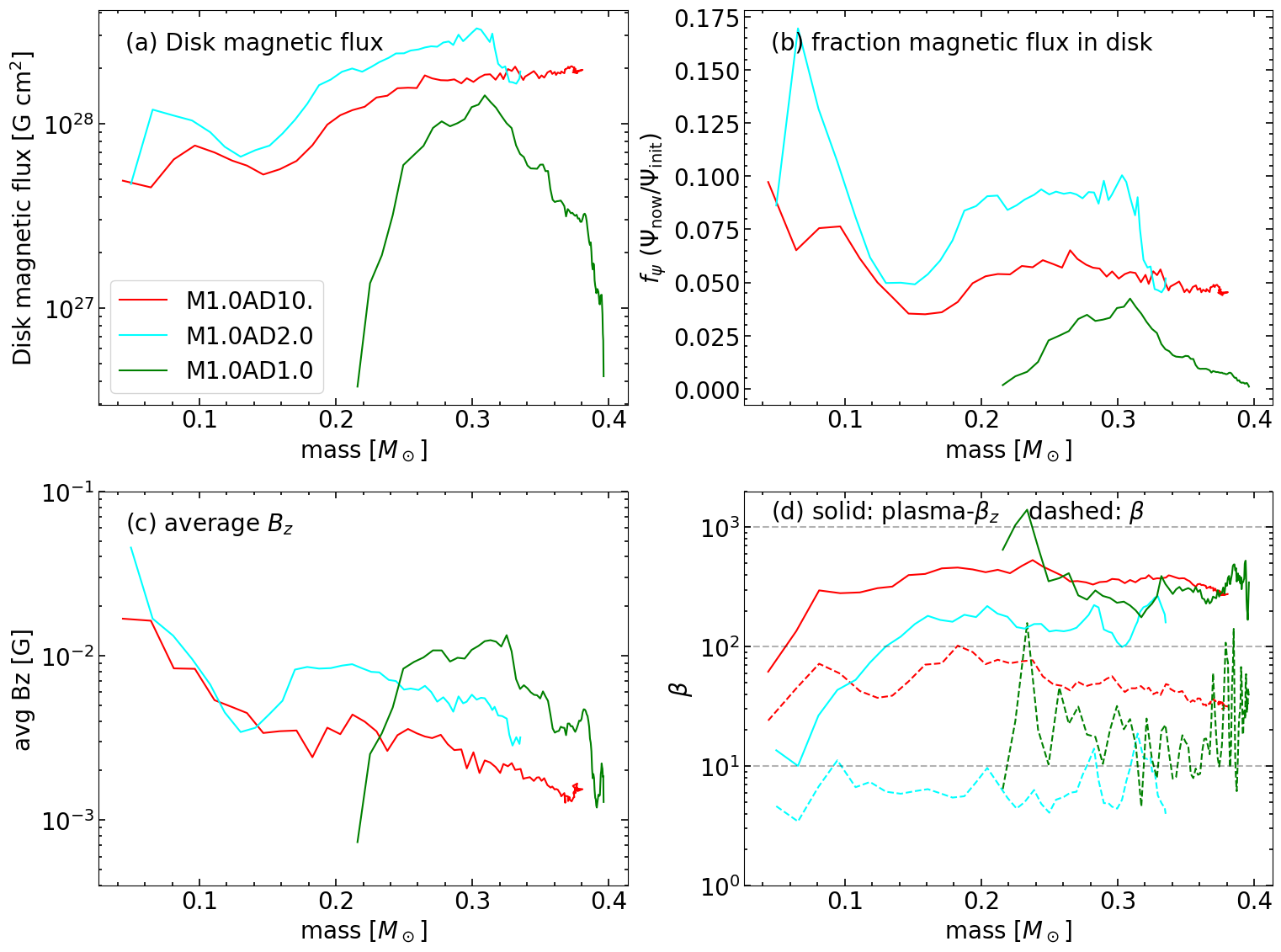}
    \caption{Evolution of the disk magnetic field as a function of the combined star and disk mass $M_{d*}$. 
    Plotted are (a) the vertical magnetic flux threading the disk midplane vertically $\Psi_{z,mid}$; (b) the disk magnetic flux fraction $f_\Psi$ defined in equation~(\ref{equ:fPsi}) of the text; (c) the vertical field strength averaged over the disk midplane, and (d) two versions of plasma-$\beta$, $\beta_z$ (solid lines) and $\beta$ (dashed), averaged over the disk midplane. 
    }
    \label{fig:DiskFieldEvol}
\end{figure*}

An important quantity characterizing the degree of global disk magnetization is the total magnetic flux threading vertically through the disk midplane $\Psi_{d,mid}$. Its evolution as a function of the combined star and disk mass $M_{d*}$ is shown in panel (a) of Fig.~\ref{fig:DiskFieldEvol}. After some initial adjustment, there is a general trend of increasing disk magnetic flux with increasing mass for the star-disk system; the decline in the reference model M1.0AD1.0 after $M_{d*}\approx 0.3$~M$_\odot$ is primarily caused by a shrinking disk at later times. This general trend is not surprising because more magnetic flux is expected to be dragged into the disk as mass is added to the star-disk system. As an approximate measure of the fraction of the magnetic flux initially associated with the mass of the star and disk that remains trapped on the disk, we define a dimensionless quantity:
\begin{equation}
    f_\Psi\equiv \frac{\Psi_{d,mid}}{\Psi_{i,M}},
    \label{equ:fPsi}
\end{equation} 
where $\Psi_{i,M}$ is the magnetic flux threading a sphere enclosing the same amount of mass as $M_{d*}$ but at the beginning of the simulation ($t=0$). From panel (b), we see that the flux fraction $f_\Psi$ ranges from a few percent for the reference model M1.0AD1.0 to about $10\%$ for the intermediate magnetic diffusivity model M1.0AD2.0. Interestingly, the most diffusivity model M1.0AD10 has a flux fraction $f_\Psi$ between the two less diffusive models, with a value of approximately $5\%$. 

The non-monotonic variation of the fraction of the magnetic flux the disk inherits from the envelope, $f_\Psi$, with respect to the magnetic diffusivity is shaped by two competing effects. On the one hand, the disk size (and thus surface area) increases with increasing magnetic diffusivity, which tends to increase the trapped magnetic flux $\Psi_{z,mid}$. On the other hand, the magnetic field tends to be weaker for a larger diffusivity,  lowering the flux. The latter trend is shown explicitly in panel (c), where the vertical field strength averaged over the disk mid-plane $\langle B_{z,mid}\rangle$ is plotted. The averaged net vertical field tends to be weaker for a more magnetically diffusive model, which makes intuitive sense. There is a tendency for $\langle B_{z,mid}\rangle$ to decrease with time, with typical values of order 1-10 mG, which may be detectable through ALMA Zeeman observations \citep{Vlemmings2019, Mazzei2020, Harrison2021}, especially since the disk-averaged toroidal field is significantly stronger than the vertical component.

Interestingly, the late-time decrease in $\langle B_{z,mid}\rangle$ does not translate to a corresponding increase in the mid-plane plasma-$\beta_z$ based on the vertical field component, as shown in panel (d) of Fig.~\ref{fig:DiskFieldEvol}. Clearly, $\beta_z$ plateaus around $\sim 300$ for both the most and least diffusive models (M1.0AD10 and M1.0AD1.0 of the three shown), with that for the intermediate magnetic diffusivity model (M1.0AD2.0) somewhat lower, but still within a factor of 2. The plasma-$\beta$ based on the total field is significantly lower, reaching values of order 
$10$ for the two less diffusive models (M1.0AD1.0 and M1.0AD2.0), although there is significant time variation, particularly for the more magnetically coupled case of the two (M1.0AD1.0), indicating a toroidal domination of the disk magnetic field. The total plasma-$\beta$ for the most diffusive model stays around $\sim 40$ at late times, which is higher than the less diffusive cases but still about an order of magnitude below its corresponding $\beta_z$, indicating that the disk magnetic field is toroidally dominated even in this most diffusive case.  

To summarize, the vast majority (of order $90\%$ or more) of the magnetic flux originally threading the mass that has gone into the star and disk is stripped away from the mass, presumably through the physical mechanism of ambipolar diffusion included in the simulation. However, some contribution from the difficult-to-quantify numerical reconnection cannot be ruled out\footnote{Gas flowing along magnetic field directly connected to the disk can also lead to a mass increase in the disk without a corresponding increase in field strength, thus contributing to the disk demagnetization. The contribution is expected to be limited because the fraction of the envelope material magnetically connected to the disk is relatively small at any given time.}.
This demagnetization of the disk and stellar material is likely facilitated by the majority of the envelope material being forced into thin gravo-magneto-sheetlets with highly pinched field lines, which increases the magnetic forces and, thus, the ion-neutral drift that separates the magnetic field relative to the matter. Nevertheless, a small but highly significant fraction of the original magnetic field (up to $\sim 10\%$) is brought into the disk, enabling the disk to remain significantly magnetized (with $\beta_z$ of a few hundred). The inherited magnetic field plays a key role in the subsequent disk dynamics and evolution, as shown in \S~\ref{sec:DiskOnly} below.

\subsection{Magnetic Collapse-Retarding Force and Rotating-Braking Torque}
\label{sec:mag_forces}

\begin{figure*}
    \centering
    \includegraphics[width=\textwidth]{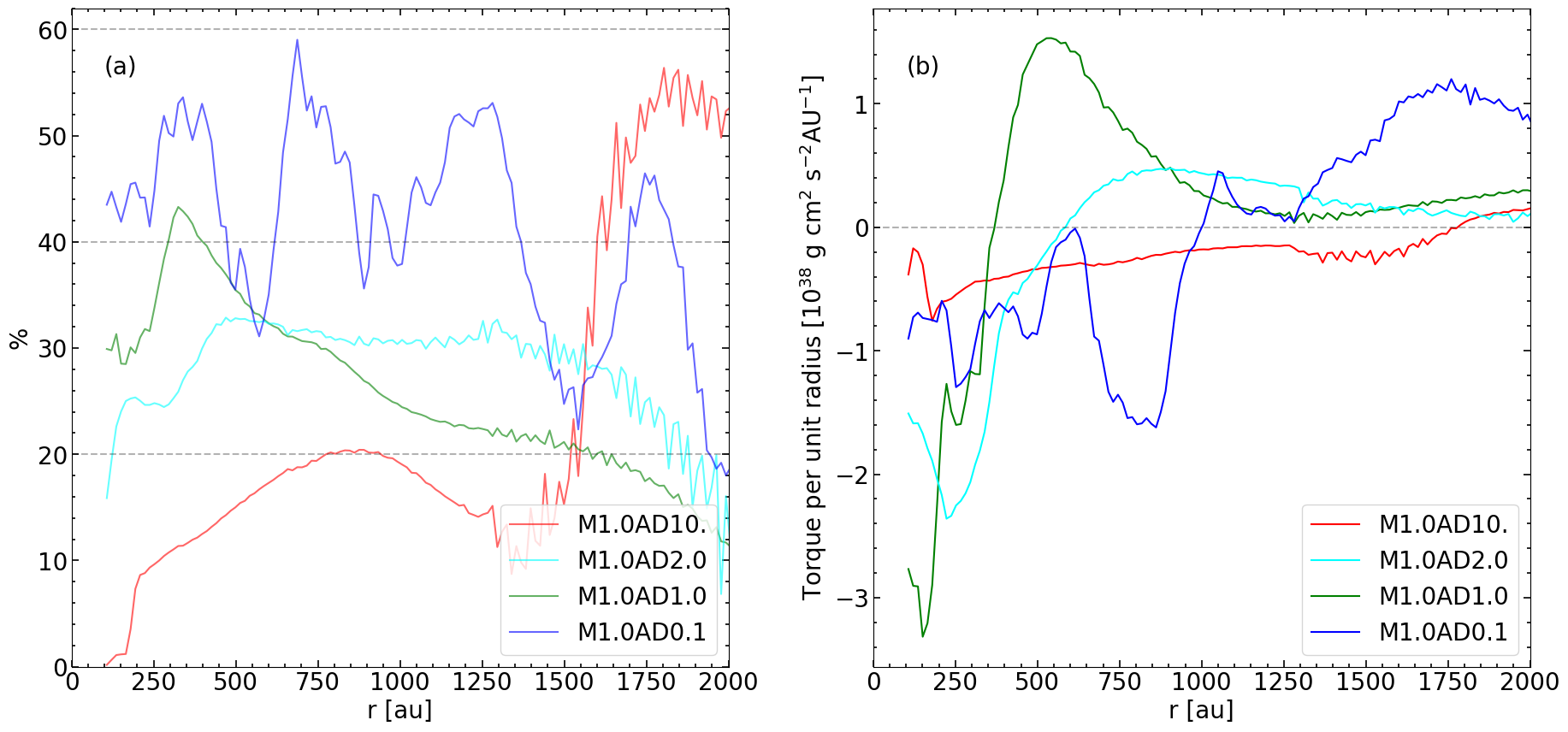}
    \caption{Plotted as a function of radius are (a) the ratio of the spherically averaged radial components of the magnetic and gravitational forces and (b) the $z-$component of the magnetic torque per unit radius for the four turbulent models with different magnetic diffusivities at the representative time. 
    }
    \label{fig:magnetic_force_r_and_phi}
\end{figure*}

We have already established in \S~\ref{sec:envelope} that most of the volume of the inner envelope is magnetically dominated, with a plasma-$\beta$ well below unity, especially in the low-density background (see Fig.~\ref{fig:slices}c and Fig.~\ref{fig:back_bones}c), and that the field is strong enough to significantly retard the envelope collapse (see Fig.~\ref{fig:slices}d and Fig.~\ref{fig:back_bones}d) and regulate the disk formation through magnetic braking (see Fig.~\ref{fig:col_den}). In this subsection, we quantify the radial magnetic force that retards the gravitational collapse and the magnetic torque from the toroidal magnetic force that redistributes the angular momentum. The radial magnetic force density is calculated with
\begin{equation}
    F_{B,r} = \mathbfit{F}_B\cdot\hat{r}_\mathrm{sph} = \Big[\frac{1}{4\pi}(\nabla\times \mathbfit{B})\times \mathbfit{B}\Big] \cdot \hat{r}_\mathrm{sph}
    \label{eq:Mag_ForceDensity}
\end{equation}
and the $z$-component of the magnetic torque density is defined as
\begin{equation}
    \tau_B = \mathbfit{F}_B \cdot (\mathbfit{r}_\mathrm{cyl}\times \hat{z})
    \label{eq:TorqueDensity}
\end{equation}
where $\hat{r}_\mathrm{sph}$ is a unit vector in the spherical radial direction with the star at the origin, and $\mathbfit{r}_\mathrm{cyl}$ is the radial vector in a cylindrical coordinate centered on the star, with the polar axis along the Cartesian $\hat{z}$ axis.

Panel (a) of Fig.~\ref{fig:magnetic_force_r_and_phi} plots as a function of radius the ratio of spherically averaged radial components of the magnetic and gravitational forces at the representative time. The least magnetically diffusive model M1.0AD0.1 has the strongest radially outward magnetic force, reaching values of order half of the local gravity, although there is a strong spatial variation. The more severe magnetic retardation is the reason for the slower growth in the mass of its star plus disk compared to other models (see the solid blue curve in the left panel of Fig.~\ref{fig:overview}). The radial magnetic force is less extreme for the more diffusive models but remains significant, reaching values typically between $\sim 20 - 40\%$ of the local gravity in the inner envelopes. It explains why the collapse of the inner envelope is significantly sub-free-fall (see Fig.~\ref{fig:back_bones}d).

Panel (b) of Fig.~\ref{fig:magnetic_force_r_and_phi} plots as a function of radius the $z-$component of the magnetic torque per unit radius 
\begin{equation}
    \frac{d T_B}{d r}=\int_0^{2\pi}\int_0^{\pi} \tau_B \ r^2\ \sin(\theta)\ d\theta\ d\phi
\end{equation}
where $r$, $\theta$, and $\phi$ are the spherical radius, polar angle, and azimuthal angle in a spherical polar coordinate system centered on the protostar, and $\tau_B$ the $z-$component of the magnetic torque density given in equation~(\ref{eq:TorqueDensity}). The magnetic torque is negative in the inner part of the envelope (and disk), indicating that it typically removes angular momentum from the gas at small radii. At larger distances, the magnetic torque generally becomes positive, indicating that the magnetic force acts in the positive azimuthal direction to spin up the gas. The sign change of the torque is a natural consequence of the outward magnetic transport of angular momentum in a collapsing envelope that is magnetized and rotates differentially, with the inner part rotating faster than the outer part on average.
The transport is least efficient in the most diffusive model M1.0AD10, enabling a large, massive, persistent disk to form. As the magnetic diffusivity decreases from model M1.0AD10 to M1.0AD2.0 to M1.0AD1.0, the magnetic torque increases, leading to stronger inner envelope braking and forming a smaller, less massive, and more transient disk. For the least diffusive model M1.0AD0.1, there is so little angular momentum left in the inner envelope that a relatively small magnetic torque is enough to transport it outward, leading to a tiny, transient disk. 

\subsection{Sheetlet-Fed Envelope Accretion onto Disk}
\label{sec:accretion}

\begin{figure*}
    \centering
    \includegraphics[width=\textwidth]{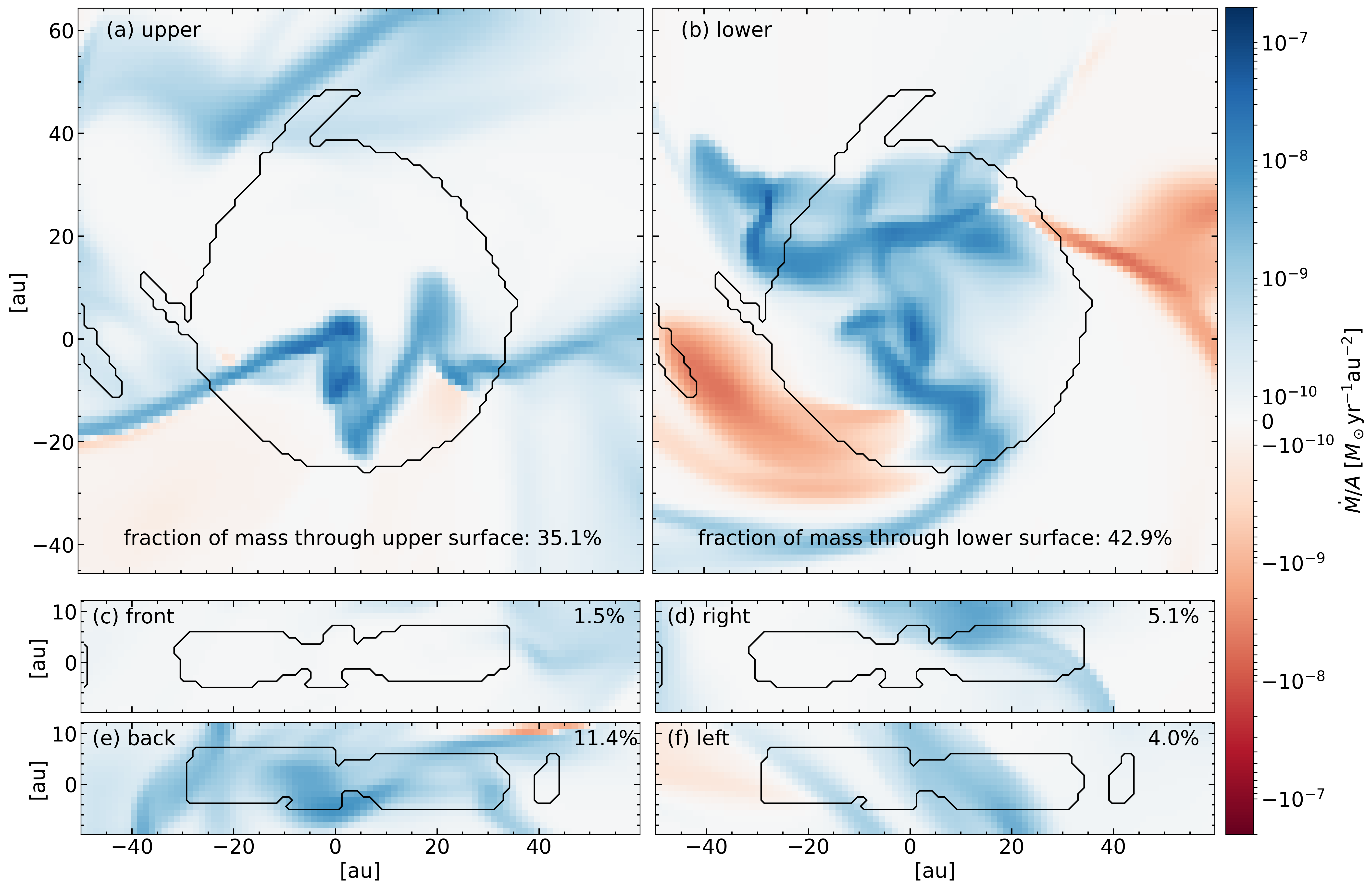}
    \caption{Highly anisotropic envelope mass accretion onto the disk. Plotted are the mass accretion rate per unit area through the six surfaces of a rectangular box enclosing the disk for the reference model at the representative time. Accretion into the box/disk is colored blue and out of the box/disk brown. 
    }
    \label{fig:disk_acc_through_box}
\end{figure*}

We have established in \S~\ref{sec:envelope} that there are two types of regions in the inner envelope -- the gravity-dominated dense gravo-magneto-sheetlets and the magnetically dominated low-density background. The background occupies most of the volume, but the sheetlets dominate the mass and mass accretion rate. This bimodal behavior is expected to have an impact on how mass is fed from the inner envelope onto the disk, as we demonstrate in Fig.~\ref{fig:disk_acc_through_box}. 

The figure shows the mass and angular momentum accretion rates across the six surfaces of a flattened rectangular box enclosing the disk for the reference model at a representative time. Two features are worth noting. Firstly, a large majority of the mass ($\sim 78\%$) lands on the disk through the upper and lower surfaces rather than the sides. Secondly, the mass accretion onto the upper and lower disk surfaces is highly inhomogeneous, with most of the mass flux concentrated into narrow filaments that span a range of disk radii. Both features are direct consequences of the dense gravo-magneto-sheetlets being warped out of the equatorial plane of the system by turbulence (see Fig.~\ref{fig:3D} and Fig.~\ref{fig:slices}); these out-of-the-plane collapsing sheetlets tend to feed the disk through its upper and lower surfaces rather than equatorially through its outer edge.

The mass accretion primarily through the upper and lower disk surfaces has implications on the net magnetic flux $\Psi_z$ threading the disk. As shown in Fig.~\ref{fig:DiskFieldEvol}b, the disk is drastically demagnetized relatively to the simplest expectation, retaining only a small fraction ($\sim 10\%$ or less) of the original magnetic flux $\Psi_{i,M}$ initially threading the mass $M$ that has gone into the star and disk. Part of this demagnetization occurs in the sheetlets (which are primarily responsible for feeding the disk) before they approach the disk as a result of the ambipolar diffusion, which allows the sheetlet material to collapse without dragging the (highly pinched) field lines along with it at the same speed; the same outward-directing magnetic (tension) force that significantly retards the gravitational collapse of the sheetlets also pushes the field lines to diffuse outward relative to the collapsing material through ambipolar diffusion. The demagnetization is likely helped by the disk mass accretion through the upper and lower surfaces, since vertical accretion through a horizontal plane can add mass to the disk, but not the {\it net} vertical magnetic flux $\Psi_z$; the latter is increased primarily through horizontal mass accretion near the equatorial plane, which is greatly reduced by the strong warping of the sheetlets. 
The dense sheetlets are the primary conduits for supplying the disk with not only mass but also angular momentum and thus are expected to affect the disk dynamics and evolution. In particular, the specific angular momentum of the sheetlet material landing on the disk may not match that of the Keplerian disk at the landing site. We will discuss this and other aspects of this new mode of disk-feeding in \S~\ref{sec:DiskOnly} below. 

\section{Protostellar Disk Structure and Evolution}
\label{sec:DiskOnly}

In this section, we examine more closely the structure and evolution of the disks formed in our simulations, focusing on the roles played by gravity, magnetic fields, and feeding by the highly structured inner envelope in the three models with persistent disks (M1.0AD10., M1.0AD2.0, and M1.0AD1.0; see the top row of Fig.~\ref{fig:toomerQ}). 

\subsection{Are Early Disks Gravitational Stable?}
\label{sec:disk_gravity_stability}

\begin{figure*}
    \centering
    \includegraphics[width=\textwidth]{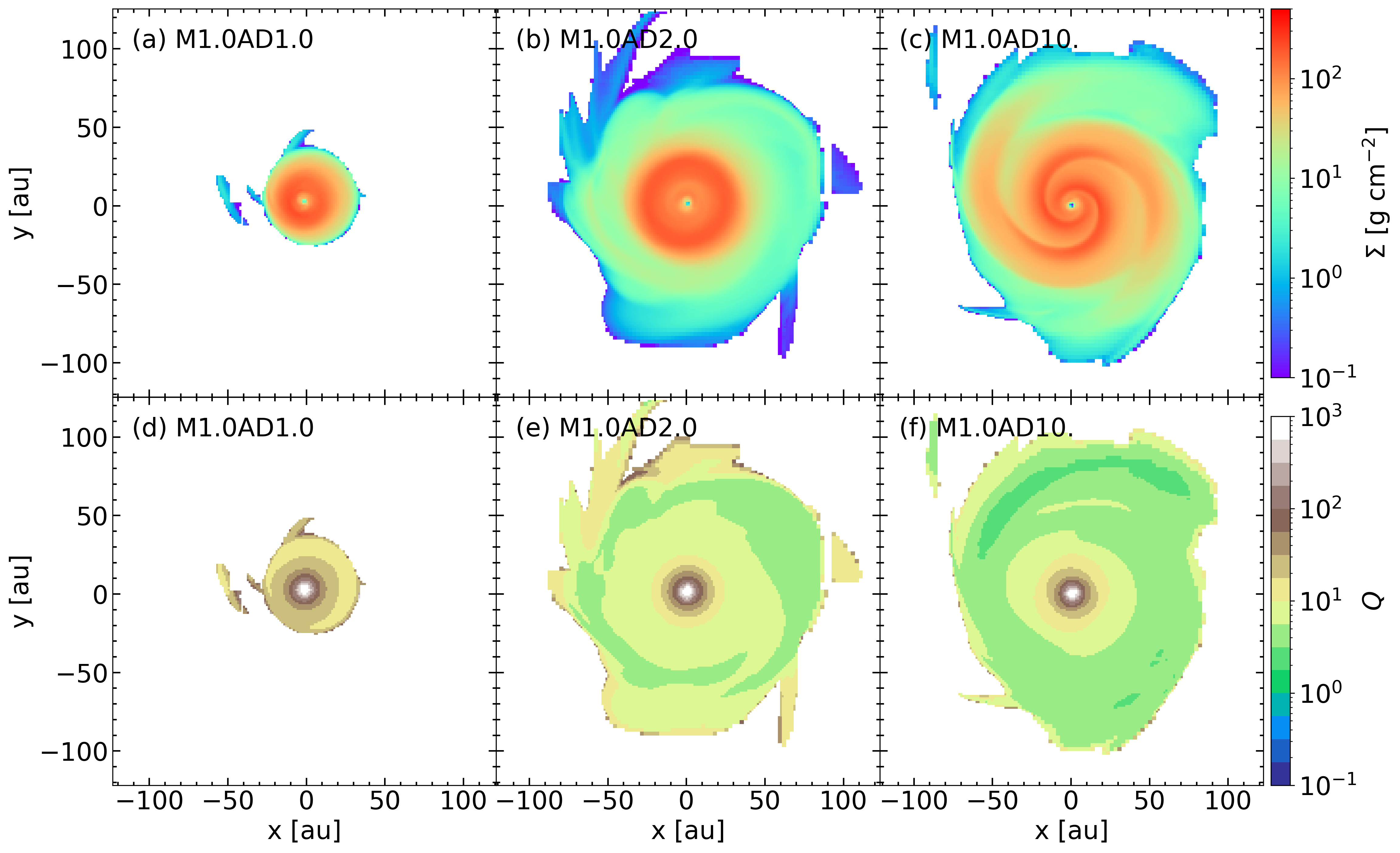}
    \caption{Gravitational stability of disks in formation. Plotted are the column density maps (top row) and the distributions of the Toomre Q parameter for disk gravitational stability (bottom row) for models M1.0AD1.0, M1.0AD2.0, and M1.0AD10 at the representative time. }
    \label{fig:toomerQ}
\end{figure*}

We start by addressing whether all early disks are gravitationally unstable. It would be the case if the magnetic field needs to decouple from the matter to such a degree in order to form the disk in the first place that, once a disk has formed, the field becomes too weak and/or too decoupled to the gas to drive the subsequent disk evolution. In such a case, the disk would accumulate mass until it becomes gravitationally unstable, forming spiral arms to redistribute angular momentum and drive disk accretion. Prominent, persistent spiral arms are indeed present in the most diffusive model M1.0AD10 (see Fig.~\ref{fig:toomerQ}c), indicating that the disk formed in this case was gravitationally unstable. This conclusion is strengthened by Fig.~\ref{fig:toomerQ}, which plots the Toomre parameter: 
\begin{equation}
    Q = \frac{c_s\Omega}{\pi G\Sigma}
\end{equation}
where $c_s$ is the isothermal sound speed, $\Omega$ the local Keplerian angular speed, and $G$ and $\Sigma$ the gravitational constant and disk column density respectively. It is clear from panel (f) that $Q$ is larger than, but close to, unity in the outer half of the disk, where the spirals are the most prominent. Evidently, the non-linear development of the spirals has driven the disk towards a marginally stable state. 
%
%
In the less magnetically diffusive model M1.0AD2.0, the $Q$ parameter remains above $\sim 4$ everywhere (see panel [e]), indicating the disk is largely gravitational stable at the time shown. However, its outer edge occasionally approaches closer to the gravitational instability, with transient spiral arms developing and disappearing.
In contrast, the least magnetic diffusive model of the three, M1.0AD1.0, has a $Q$ parameter of order $10$ or more nearly everywhere on the disk (panel d), indicating that the disk formed in this case is gravitationally stable. This is consistent with the surface density distribution, which shows only transient, weak spirals that likely arise from the rotational smearing of patchy materials landed on the disk. We conclude that not all early disks are gravitationally unstable, and whether they become so depends on the magnetic diffusivity. Instability is more likely to develop in more magnetic diffusive cases. It makes physical sense because the disks are bigger and more massive in the more magnetically diffusive cases, as shown in Fig.~\ref{fig:overview}.

The fact that some disks become gravitationally unstable while others do not naturally raises the question: which mechanism is more important in driving disk angular momentum evolution:  gravitational torque or magnetic braking? We will address this question next. 

\begin{figure}
    \centering
    \includegraphics[width=0.5\textwidth]{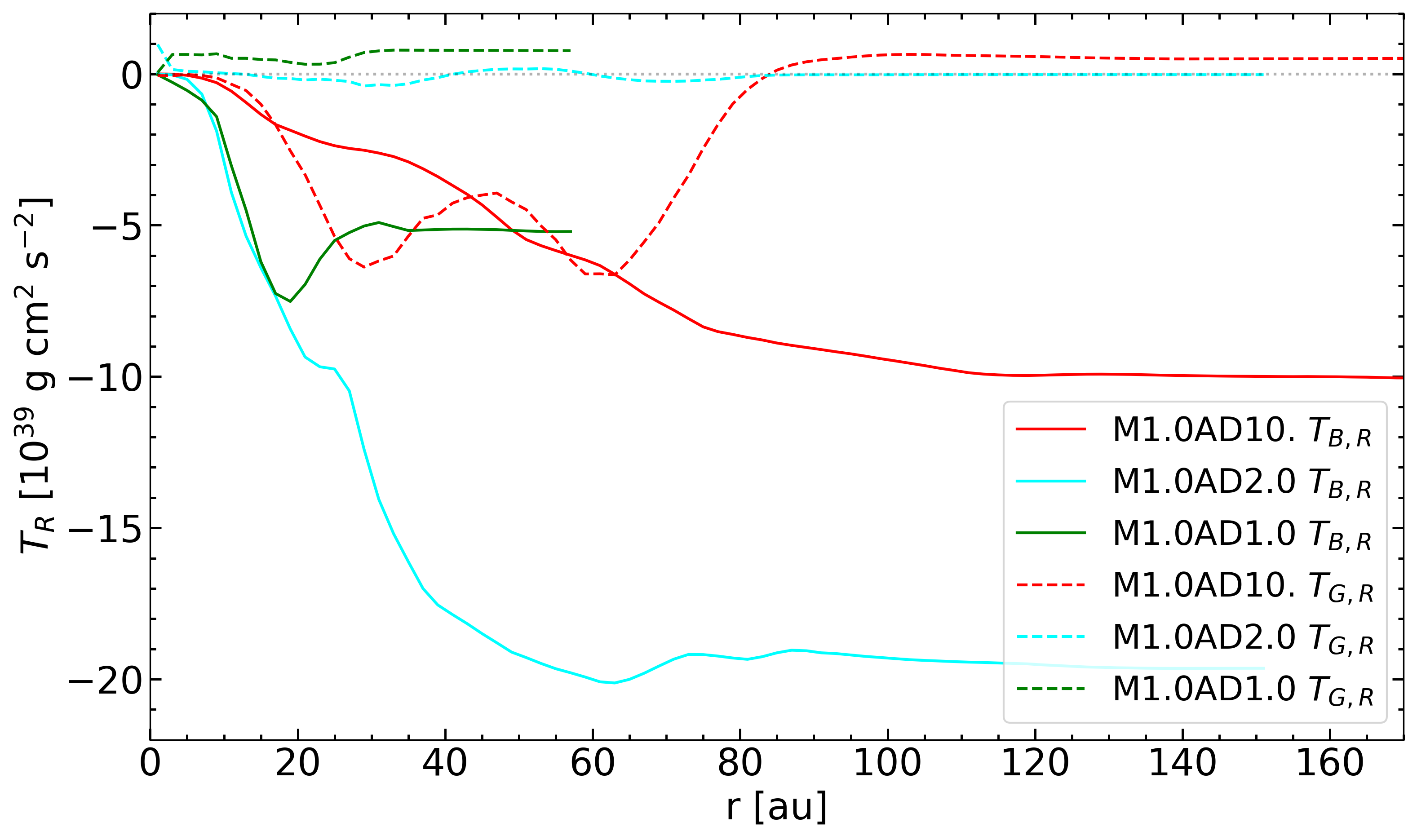}
    \caption{Cumulative magnetic (solid lines) and gravitational (dashed) torques inside the disk up to a given cylindrical radius $R$ at the representative time.}
    \label{fig:BandG_force_on_disk}
\end{figure}

\subsection{Magnetically or Gravitational Driven Angular Momentum Evolution?}
\label{sec:disk_ang_mom_transport}

To quantify the roles of gravity and magnetic fields in disk angular momentum evolution, we compute the azimuthal components of the gravitational and magnetic force densities, $F_{G,\phi}$ and
$F_{B,\phi}$, everywhere inside the disk, multiply them by the cylindrical radius $r_{\rm cyl}$ to obtain the local gravitational and magnetic torque densities, $\tau_G$ and $\tau_B$, and then sum all torques up to a given cylindrical radius $R$ to obtain cumulative gravitational and magnetic torques, $T_{G,R}$ and $T_{B,R}$, as a function of the cylindrical radius $R$:
\begin{equation}
    T_{G, R} = \int_{z\in\mathrm{disk}}\int_{\phi\in\mathrm{disk}}\int_{r_{\rm cyl}=0}^{r_{\rm cyl}=R} F_{G,\phi}\ r_{\rm cyl} dV
\end{equation}
\begin{equation}
    T_{B, R} = \int_{z\in\mathrm{disk}}\int_{\phi\in\mathrm{disk}}\int_{r_{\rm cyl}=0}^{r_{\rm cyl}=R} F_{B,\phi}\ r_{\rm cyl} dV
\end{equation}
where $F_{G,\phi}$ and $F_{B,\phi}$ are the azimuthal component of the gravitational and magnetic force, respectively, $r_{\rm cyl}$ the cylindrical radius from the $z-$axis, $dV$ the infinitesimal volume element in cylindrical coordinates.
The cumulative torques are shown in Fig.~\ref{fig:BandG_force_on_disk}. 

We will start the torque discussion with the most diffusive model M1.0AD10, where prominent spirals are present, so the gravitational torque is expected to be significant. In this case, the cumulative gravitational torque $T_{G,R}$ is plotted as the dashed red line in Fig.~\ref{fig:BandG_force_on_disk}. In the inner part of the disk, $T_{G,R}$ decreases with radius $R$ monotonically, reaching a (negative) minimum around a radius $R \sim 30$~au, indicating that the net azimuthal gravitational force in this region is negative, acting to brake the disk rotation and remove its angular momentum. Part of the removed angular momentum is transported outward to the adjacent region between $\sim 30$ and $\sim 45$~au, where $T_{G,R}$ increases to a local maximum. This region's net azimuthal gravitational force is positive, speeding up the disk rotation and increasing its angular momentum. As the radius increases, the pattern repeats itself, with gravity first removing angular momentum from the disk before giving it back to the disk. By the outer edge of the disk at $R\sim 100$~au, nearly all of the angular momentum extracted from the disk material by the gravitational torque is returned to the disk, which makes sense because the gravitational torque is mostly internal to the disk and, thus, can redistribute angular momentum between different parts of the disk but not efficiently remove angular momentum from the disk as a whole. This behavior contrasts with the magnetic torque, shown as the solid red line in Fig.~\ref{fig:BandG_force_on_disk}. Although small fluctuations exist, there is a general trend for the cumulative magnetic torque $T_{B,R}$ to decrease (becoming more negative) with radius, indicating that the net azimuthal magnetic force is predominantly negative and acts to brake the disk rotation. Most importantly, at large radii, $T_{B,R}$ plateaus to a value more negative than the minimum (negative) $T_{G,R}$, indicating that the magnetic torque is important even in this most magnetically diffusive case: it removes angular momentum from the disk at a rate comparable to or larger than the rate with which the gravitational torque redistributes angular momentum from one part of the disk to another. 

In the less magnetically diffusive model M1.0AD2.0 where a relatively large, persistent disk is also formed, the gravitational torque acts to redistribute angular momentum within the disk as well, but at a rate far below that of the magnetic torque (compare the dashed and solid cyan curves in Fig.~\ref{fig:BandG_force_on_disk}). Indeed, the magnetic torque removes angular momentum from the disk as a whole at a rate significantly higher than that in the more diffusive model M1.0AD10, likely because its disk is more strongly magnetized (see Fig.~\ref{fig:DiskFieldEvol}c,d). Therefore, in this case, the magnetic field completely dominates gravity in controlling the disk's angular momentum evolution. The same is true for the least magnetically diffusive, reference model M1.0AD1.0, where the cumulative magnetic torque $T_{B,R}$ is much larger than the cumulative gravitational torque $T_{G,R}$ in magnitude (compare the dashed and solid green curves in Fig.~\ref{fig:BandG_force_on_disk}). Interestingly, $T_{B,R}$ increases with the radius (becoming less negative) near the outer edge of the disk (between $\sim 20$ and $\sim 30$~au), indicating that a (relatively small) fraction of the angular momentum extracted magnetically from the inner part of the disk is returned to the outer part. However, there is the possibility that the disk's outer edge is spun up externally by a faster-rotating inner envelope magnetically connected to it. 
We conclude that enough magnetic flux is carried into the forming disk for the field to play a dominant role in regulating its angular momentum evolution. 


\subsection{Kinematically Perturbed Early Disks}
\label{sec:disk_vr}

\begin{figure*}
    \centering
    \includegraphics[width=\textwidth]{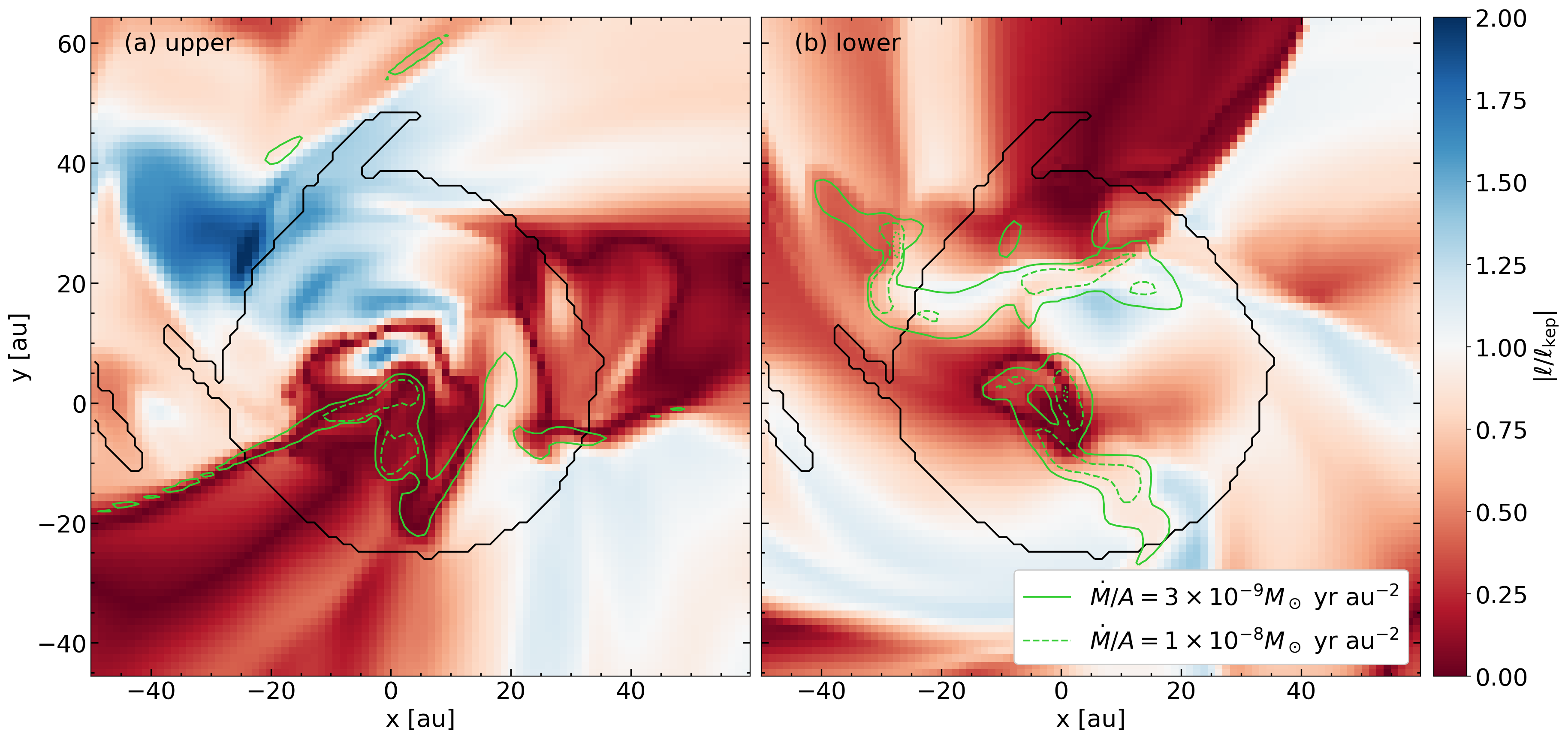}
    \caption{The ratio between the specific angular momentum carried by the infalling envelope material above (panel a) and below (panel b) the disk and that required for staying in a Keplerian orbit at the projected cylindrical radii on the disk midplane. The black contours in both panels are the projected disk midplane outline. The green curves are contours of mass feeding rate per unit area, highlighting where most envelope material lands on the disk surfaces. 
    }
        \label{fig:angmom_mismatch}
\end{figure*}

Besides the gravitational and magnetic torques discussed in the last section, accretion from the envelope can also significantly affect the disk structure and evolution. It is especially true when the accretion is highly anisotropic, as expected for an inner envelope dominated by dense sheetlets (see, e.g., Fig.~\ref{fig:3D}). Since the sheetlets can be strongly magnetically braked, there is no reason for their specific angular momenta to match those of their landing sites on the disk. The mismatch is illustrated in Fig.~\ref{fig:angmom_mismatch}, which plots the specific angular momentum above and below the disk normalized by that needed for the rotational support against gravity on the disk midplane for the reference model at the representative time. In this case, most of the mass falls onto the disk with a specific angular momentum well below the local Keplerian value (particularly for the upper disk surface), indicating that substantial mixing with the material already in the disk is required to fully incorporate the newly accreted low-angular-momentum material (which is already magnetically braked before entering the disk) into the Keplerian disk. 
As a result, the mixed material would have a lower angular momentum than the local Keplerian value and move inward. Therefore, magnetic braking of the inner envelope material at relatively large distances, particularly the dense sheetlets, provides another mechanism to drive the disk accretion by supplying the formed disk with lower angular momentum material.

\begin{figure*}
    \includegraphics[width=\textwidth]{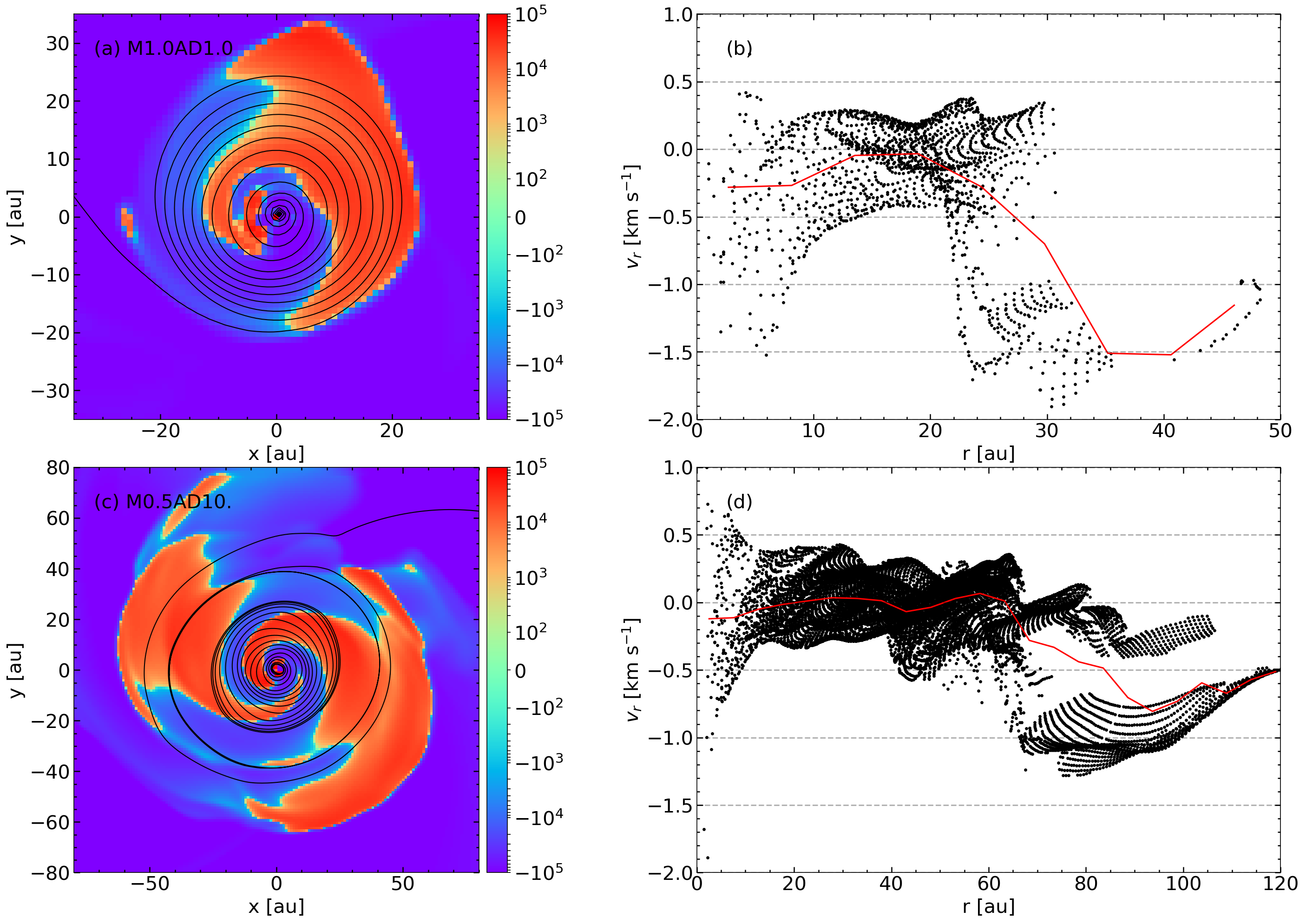}
    \caption{Perturbed disk kinematics for models M1.0AD1.0 (upper panels) and M1.0AD10 (lower) at a relatively early time when $M_{d*}=0.2$~$M_\odot$. {\bf Panels (a) and (c)}: gas radial velocity on the disk midplane. The black line in each panel is a representative flow streamline. {\bf Panels (b) and (d)}: Statistics of the gas radial velocity in the disk. A black dot represents each cell in the disk. The red lines are the averaged gas radial velocity at each radius. 
    }
        \label{fig:spiral_pattern}
\end{figure*}

Given the angular momentum mismatch discussed above and the generally patchy and episodic nature of the sheetlet-dominated disk-feeding, deviations from a perfectly circular Keplerian disk are to be expected. The deviation from a circular orbit is illustrated in Fig.~\ref{fig:spiral_pattern}a, which plots a representative streamline on the disk midplane for the reference model at a relatively early time when the combined stellar and disk mass is $M_{d*}=0.2$~$M_\odot$. The streamline loops are clearly displaced to the upper side of the protostar and are bunched up closer together in the lower left part of the disk compared to other parts. Interestingly, the trajectory spirals inwards quickly, shrinking in radius by a factor of two in only half a dozen turns. The rapid shrinkage is indicative of a significant radial velocity component, which is indeed the case, as shown by the color map in panel (a) and scatter diagram in panel (b) of Fig.~\ref{fig:spiral_pattern}. Although some regions move inward (the reddish regions in panel [a]) and others outward (purple regions), the inward motions tend to be faster than the outward ones, reaching values as high as $\sim 0.5$~km/s or more, comparable to the local sound speed. The expanding and contracting regions form a ``yin-yang" pattern. It is related to the non-circular orbits highlighted by the streamline. We note that the speeds and distributions of expanding and contracting motions are variable in time, 
which is not surprising in this (reference) case given that the disk remains significantly magnetized (see Fig.~\ref{fig:Bz_map}) and strongly braked (see Fig.~\ref{fig:BandG_force_on_disk}) by both non-uniform magnetic fields and patchy low-angular momentum material accreted from the envelope (see Fig.~\ref{fig:angmom_mismatch}). 

The disks formed in the more diffusive models contract more slowly on average. It is illustrated in Fig.~\ref{fig:spiral_pattern}d, which plots the radial component of the velocity in each cell of the disk midplane in the most diffusive model M1.0AD10. Although substantial infall exists near the outer edge, the number of cells with expanding and contracting motions are comparable inside a radius of $\sim 60$~au, leading to little net orbital shrinkage, in agreement with the outermost streamline shown in panel (c). A prominent ``yin-yang" pattern also exists in the radial velocity map, but, unlike the reference case discussed above, it is primarily caused by the spiral arms in the disk surface density, which produces non-circular orbits with misaligned semi-major-axes, as illustrated by the streamlines in the central region. 
The spiral arms are strong enough to cause significant derivations from the local Keplerian rotation speed (of order 0.2 to 0.8 times the sound speed), which may be detectable with high spatial and spectral resolution ALMA observations.
In this case, the gravitationally induced kinematic perturbations dominate those from magnetic braking and patchy envelope accretion. 

\section{Discussion}
\label{sec:discussion}


\subsection{Towards a Picture of Protostellar Envelope Dominated by Gravo-Magneto-Sheetlets}
\label{sec:Sheetlet-Streamer}



As stressed in \S~\ref{sec:intro}, the gravitational collapse of a smooth molecular cloud core threaded by a large-scale, dynamically significant magnetic field naturally leads to a dense, collapsing, equatorial sheet -- the classical pseudodisk \citep{Galli&Shu1993} because the magnetic resistance to gravity is intrinsically anisotropic. Turbulence or other perturbations, such as clumpiness, in the star-forming core tend to distort the pseudodisk into dense, collapsing, gravo-magneto-sheetlets, which appear as dense filaments in mass density or column density maps (see, e.g., Fig.~\ref{fig:overview}). Such filaments have been reported in many core-collapse simulations with both magnetic fields and turbulence (e.g., \citealp{Santos-Lima2012, Seifried2013, Li2014, Seifried2015_turb, Kuffmeier_2017, Matsumoto2017, Gray2018, Lam2019, Wurster2020_I}, \citealp{Mignon-Risse2021}; see \citealp{Tsukamoto2022PPVII} for a review). The prevalence of gravo-magneto-sheetlets in numerical simulations, coupled with their simple physical origin, adds weight to the notion that they are an integral part of protostellar envelopes. This emerging picture has implications both observationally and theoretically. 

On the observation side, it is tempting to identify the collapsing gravo-magneto-sheetlets with the so-called ``accretion streamers" observed around an increasing number of protostars, including, e.g., the Class 0 sources VLA 1623 \citep{Hsieh2020}, IRAS 03292
+3039 \citep{Pineda2020}, IRAS 16293-2422 \citep{Murillo2022}, and Lupus 3-MMS \citep{Thieme2022}, and Class I sources L1498 \citep{Yen2014}, HL Tau \citep{Yen2019}, and Per emb-50 \citep{Valdivia-Mena2022}. To strengthen the identification, additional information is needed. In particular, it would be interesting to determine using, e.g., ALMA, whether the accretion streamers are associated with highly pinched magnetic field lines \citep[e.g.][]{Sanhueza2021}, which is a strong prediction of the picture (see Fig.~\ref{fig:slices}a). 

Another prediction of the picture is that the inner envelope collapses at a speed substantially below the free-fall speed (see Figs.~\ref{fig:slices}d and \ref{fig:back_bones}d) because of strong magnetic retardation to the gravitational collapse, particularly in the dense gravo-magneto-sheetlets. The sub-free-fall collapse has been inferred around a few protostars, including, e.g., L1527 IRS \citep{Ohashi2014} and TMC-1A \citep{Aso2015}, which could be due to magnetic support. It remains to be determined whether this is generally true. A difficulty is that an accurate determination of the stellar mass is needed to ascertain the free-fall speed, which is not easy for deeply embedded protostars. Ultimately, the most direct observational test of the picture is a direct measurement of the magnetic field strength in the envelope through the Zeeman effect. Its feasibility with ALMA 
will be explored in a separate investigation (Mazzei, R. et al. 2023, under review). 


On the theory side, the domination by dense gravo-magneto-sheetlets has implications on several aspects of the protostellar envelope, including chemistry and grain growth. Since the sheetlets are much denser than their surrounding background, chemical reactions and freeze-out of molecules onto dust grains should occur faster there than naively expected for 
a non-clumpy envelope. Similarly, grain growth in the envelope should be enhanced because the higher density shortens the grain-grain collision time, and the magnetic retardation of collapse allows more time for the grains to grow. It could be further sped up by the sharp pinching of the field lines in the dense sheetlets, which increases the magnetic force and, thus, the relative drift (and thus collision) speed between charged and neutral grains \citep[e.g.][]{Guillet2020}. In addition, the segregation of the inner envelope into a highly structured medium with dense, thin sheetlets embedded in a magnetically dominated background should affect the detailed properties of the local turbulence, which may control the grain growth rate \citep[e.g.][]{Ormel2009, Tu2022}. It would be interesting to quantify these effects in the future. 



\subsection{Magnetically Driven Evolution of Heterogeneous Protostellar Disks}

A direct consequence of the picture of protostellar envelope dominated by gravo-magneto-sheetlets is that the protostellar disk is fed primarily from above and below the equatorial plane (see, e.g., Fig.~\ref{fig:disk_acc_through_box})\footnote{It is possible for a stellar or inner disk wind (not included in the present model) to clear out the collapsing streams above and below the disk, forcing most of the envelope accretion to the equatorial plane, especially at relatively late times when the outflow cavity has a large opening angle near the base.}.  Such patchy off-the-equatorial plane accretion has been found in other simulations \citep[e.g.][]{Lee2021} and is more prevalent in more turbulent and better magnetically coupled envelopes. It is expected to produce heterogeneities on the disk that could evolve into substructures such as transient spirals as well as rings and gaps \citep[e.g.][]{Kuznetsova2022}. However, the relatively low spatial resolution on the disk scale does not allow us to study the substructure formation properly in our global core collapse and disk formation simulations. Another indication of the protostellar disks being more dynamically active than the more evolved protoplanetary disks is the large deviation from pure Keplerian rotation found in our simulations (see Fig.~\ref{fig:spiral_pattern}). 


As mentioned in \S~\ref{sec:intro} and stressed by \cite{Lesur2022PPVII}, a central question for the protostellar disk evolution is: how strong is the disk magnetic field? It is related to the classical magnetic flux problem for the central star, where the stellar field strength would be orders of magnitude above the observed values if all of the magnetic flux on the core scale is dragged into the star \citep{Mestel1956}. If dragged into the disk instead, it would rapidly brake the rotation, creating a corresponding ``magnetic flux problem for the disk." Therefore, a crucial task for magnetized disk formation calculations is to evaluate the fraction of magnetic flux dragged from the envelope into the disk $f_\Psi$. 

We find that the fraction $f_\Psi$ depends on the magnetic diffusivity in a complex manner. One may expect a smaller flux fraction $f_\Psi$ for a more magnetically diffusive case. However, our most magnetically diffusive model M1.0AD10 has a larger $f_\Psi$ than the reference model M1.0AD1.0 (see Fig.~\ref{fig:DiskFieldEvol}b), which is less diffusive by a factor of ten. One reason for this counter-intuitive result is that the disk is larger in the former case. Another (less obvious) reason is that the gravo-magneto-sheetlets become thinner with more sharply pinched field lines in the less diffusive case, facilitating the de-magnetization of the dense material in the sheetlets through ambipolar diffusion. 

Our simulations show that $f_\Psi$ ranges from a few to $\sim 10\%$ for the three models with persistent disks (see Fig.~\ref{fig:DiskFieldEvol}b). They yield a disk-averaged vertical field strength between $\sim 1$ to $\sim 10$~mG (see Fig.~\ref{fig:DiskFieldEvol}c), which is about one order of magnitude lower than the values found by \citet[][see their Fig.~1]{Masson2016} and \citet[][see their Fig.~2]{XuKunz2021}, as reviewed by \citet{Tsukamoto2023}\footnote{The lower average vertical field strengths in our disks compared to those of \citet{Masson2016} and \citet{XuKunz2021} may come from different ambipolar diffusivities used in our simulations compared to theirs \citep[e.g.,][]{Marchand2016}. For example, a higher diffusivity in the envelope would lower the magnetic flux delivered by the protostellar collapse to the disk region (and hence reducing the field strength there). It is also possible that the turbulence included in our simulations (but not theirs) enabled a more efficient outward magnetic flux transport relative to mass in the envelope for a given microscopic ambipolar diffusivity. 
}.
Nevertheless, the vertical field is strong enough to yield a disk-averaged plasma-$\beta_z$ based on this component alone of order a few hundred, which, as already stressed by others \citep[e.g.][]{Lam2019, Hennebelle2020, Lee2021}, is much lower than the values of order $10^4$-$10^{5}$ typically adopted for MHD simulations of evolved (Class II) protoplanetary disks. The disk-averaged plasma-$\beta$ based on the total field strength is significantly lower still, reaching values as low as $\sim 10$. 

The relatively low values of plasma-$\beta$ make it possible for the magnetic field to dominate the disk angular momentum evolution in our relatively less magnetically diffusive models M1.0AD1.0 and M1.0AD2.0. These models illustrate an important conceptual point that is not obvious a priori: it is possible to de-magnetize the collapsing envelope enough for a persistent disk to form but not so much as to render the magnetic field insignificant in driving the subsequent disk evolution. These disks are gravitational stable, with a Toomre $Q$-parameter well above unity. This conclusion is strengthened by the high-resolution non-ideal MHD simulations of \citet{Lee2021}, where a similarly gravitationally stable disk was formed  with a strong enough disk field to drive fast accretion. Even in our most magnetically diffusive model M1.0AD10, the magnetic field remains significant in removing angular momentum from the marginally gravitational unstable disk with prominent spiral arms. 

In contrast, \citet{Xu2021, XuKunz2021} found that the protostellar disks in their 3D simulations are marginally gravitationally unstable, and the magnetic field plays little role in the disk angular momentum transport despite a much higher field strength than in our simulations. We believe the inefficient magnetic braking is caused by a large magnetic diffusivity on the disk that essentially decouples the field lines from the disk material, as evidenced by the nearly vertical and uniform disk field with a much smaller toroidal component. 

We conclude that the disk can inherit enough magnetic flux from the envelope for the field to play a significant, even dominant, role in the disk angular momentum transport if it remains sufficiently well coupled to the bulk disk material. The degree of coupling depends on several uncertain factors, including the abundance of small grains, the depletion factor for metals such as Mg, and the cosmic ray ionization rate $\zeta$ \citep[e.g.][]{Lesur2021_review}. For example, $\zeta$ can be reduced by a magnetized outflow from the disk and/or central star that can severely attenuate low-energy cosmic rays \citep[e.g.][]{Cleeves2013}. On the
other hand, the cosmic ray flux can potentially be enhanced by local particle acceleration at
shocks near the stellar surface or in protostellar jets \citep{Padovani2018, Offner2019}. Ultimately, the degree of magnetic coupling must be constrained by comparing model predictions \citep[e.g.][]{Kuffmeier2020} with detailed observations of dust and gas, particularly molecular ions \citep[e.g.][]{Aikawa2021}. 
%

If the magnetic field is inefficient in transporting angular momentum, disk evolution must rely on gravitational torque from spiral arms. Although spiral arms have been detected in some protostellar disks \citep[e.g.][]{Lee_Li2020}, they are uncommon. A large optical depth at the observing wavelength or a low resolution could hinder the detection. High-resolution (0.04$^{\prime\prime}$, or $\sim 7$~au) observations are now available for 12 Class 0 and 7 Class I disks in nearby star formation regions from the ALMA Large Program eDisk \citep{Ohashi2023}. None of the disks shows obvious spirals. While it is important to quantify the extent to which the optical depth and other effects (such as the disk inclination) hinder their detection through detailed modeling and additional observations, the simplest conclusion based on the currently available data is that there is little compelling observational evidence for the majority of these early disks to have strong spirals capable of driving their fast accretion. 
%
%
We conclude that magnetic transport of angular momentum remains a viable option for driving protostellar disk evolution. It does not preclude gravitational torque from playing a significant, even dominant, role in some cases.

\section{Conclusion}
\label{sec:conclusion}

We extended the non-ideal MHD simulations by \citet{Lam2019} to study magnetized cloud core collapse and disk formation including ambipolar diffusion and turbulence. Our improved simulations incorporated adaptive mesh refinement, enabling better resolution of the disk scale. Our key findings are as follows: 

1. Turbulent magnetized protostellar envelopes are dominated by dense collapsing 3D gravo-magneto-sheetlets in their inner regions. These sheetlets, which are disrupted versions of classical sheet-like pseudodisks, are embedded in a magnetically dominant (low plasma-$\beta$) background where less dense materials flow along local magnetic field lines and accumulate in the sheetlets. The gravitational collapse of the dense sheetlets, occurring across locally pinched field lines, serves as the primary mechanism for the envelope to feed mass and angular momentum to the central disk-plus-star. Due to strong magnetic retardation, the collapse of the inner envelope is typically slower than the local free-fall rate.
%
%

2. The collapsing sheetlets primarily feed the disk through its upper and lower surfaces, contributing materials with specific angular momenta that typically deviate from Keplerian values at their landing sites on the disk. Sheetlets that experience significant magnetic braking in the envelope can supply low-angular momentum materials to the disk, thereby driving disk accretion along with gravitational torque and magnetic stresses.

3. The protostellar disk inherits a small fraction (up to 10\%) of the magnetic flux from the envelope, resulting in a disk-averaged net vertical field strength of around 1-10 mG and a stronger toroidal field. Such field strengths can potentially be measured through ALMA Zeeman observations. The plasma-$\beta_z$ based solely on the vertical field component is typically a few hundred, significantly lower than the values commonly assumed for Class II protoplanetary disks. The plasma-$\beta$ based on the total field strength is even lower, reaching values as low as approximately 10.

4. The inherited magnetic field from the envelope plays a dominant role in the evolution of disk angular momentum, facilitating the formation of gravitationally stable disks in our well-coupled models. This magnetic field's influence remains significant even in our least coupled model, as it removes angular momentum from the marginally gravitationally unstable disk at a rate comparable to or greater than that caused by spiral arms. It is thus possible to demagnetize the collapsing envelope enough to allow for the formation of a persistent disk without rendering the disk field too weak to drive subsequent evolution. This magnetically driven disk evolution is consistent with the apparent scarcity of prominent spirals capable of driving rapid accretion in deeply embedded protostellar disks. However, detecting these spirals may be hindered by optical depth and resolution effects.

5. Early disks in the active formation stage experience significant perturbations in both morphology and kinematics. These perturbations arise from the rapid inward spiraling of disk material on eccentric orbits when the magnetic field remains relatively well coupled to the disk material and from spiral arms in more magnetically diffusive cases. 

6. The dense collapsing gravo-magneto-sheetlets observed in our simulations may correspond to the "accretion streamers" observed around protostars. Additional observations are necessary to confirm whether these streamers are associated with the predicted highly pinched magnetic field and exhibit sub-free-fall speeds indicative of significant magnetic retardation.

\section*{Acknowledgements}

We thank the referee for a constructive report. Resources supporting this work were provided by the NASA High-End Computing (HEC) Program through the NASA Advanced Supercomputing (NAS) Division at Ames Research Center and the RIVANNA supercomputer at the University of Virginia. YT acknowledges support from an interdisciplinary fellowship from the University of Virginia. ZYL is supported in part by NSF AST-2308199, AST-1910106, and NASA 80NSSC20K0533. KT is supported by the Japan Society for the Promotion of Science (JSPS) KAKENHI Grant Numbers JP21H04487, JP21H04495, and JP22KK0043.

\section*{Data Availability}
The data underlying this article will be shared on reasonable request to the corresponding author.



\bibliographystyle{mnras}
\bibliography{example} 

\begin{thebibliography}{}
\makeatletter
\relax
\def\mn@urlcharsother{\let\do\@makeother \do\$\do\&\do\#\do\^\do\_\do\%\do\~}
\def\mn@doi{\begingroup\mn@urlcharsother \@ifnextchar [ {\mn@doi@}
  {\mn@doi@[]}}
\def\mn@doi@[#1]#2{\def\@tempa{#1}\ifx\@tempa\@empty \href
  {http://dx.doi.org/#2} {doi:#2}\else \href {http://dx.doi.org/#2} {#1}\fi
  \endgroup}
\def\mn@eprint#1#2{\mn@eprint@#1:#2::\@nil}
\def\mn@eprint@arXiv#1{\href {http://arxiv.org/abs/#1} {{\tt arXiv:#1}}}
\def\mn@eprint@dblp#1{\href {http://dblp.uni-trier.de/rec/bibtex/#1.xml}
  {dblp:#1}}
\def\mn@eprint@#1:#2:#3:#4\@nil{\def\@tempa {#1}\def\@tempb {#2}\def\@tempc
  {#3}\ifx \@tempc \@empty \let \@tempc \@tempb \let \@tempb \@tempa \fi \ifx
  \@tempb \@empty \def\@tempb {arXiv}\fi \@ifundefined
  {mn@eprint@\@tempb}{\@tempb:\@tempc}{\expandafter \expandafter \csname
  mn@eprint@\@tempb\endcsname \expandafter{\@tempc}}}

\bibitem[\protect\citeauthoryear{{Aikawa} et~al.,}{{Aikawa}
  et~al.}{2021}]{Aikawa2021}
{Aikawa} Y.,  et~al., 2021, \mn@doi [\apjs] {10.3847/1538-4365/ac143c}, \href
  {https://ui.adsabs.harvard.edu/abs/2021ApJS..257...13A} {257, 13}

\bibitem[\protect\citeauthoryear{{Aso} et~al.,}{{Aso} et~al.}{2015}]{Aso2015}
{Aso} Y.,  et~al., 2015, \mn@doi [\apj] {10.1088/0004-637X/812/1/27}, \href
  {https://ui.adsabs.harvard.edu/abs/2015ApJ...812...27A} {812, 27}

\bibitem[\protect\citeauthoryear{{Cleeves}, {Adams}  \& {Bergin}}{{Cleeves}
  et~al.}{2013}]{Cleeves2013}
{Cleeves} L.~I.,  {Adams} F.~C.,   {Bergin} E.~A.,  2013, \mn@doi [\apj]
  {10.1088/0004-637X/772/1/5}, \href
  {https://ui.adsabs.harvard.edu/abs/2013ApJ...772....5C} {772, 5}

\bibitem[\protect\citeauthoryear{{Draine}, {Roberge}  \& {Dalgarno}}{{Draine}
  et~al.}{1983}]{Draine1983}
{Draine} B.~T.,  {Roberge} W.~G.,   {Dalgarno} A.,  1983, \mn@doi [\apj]
  {10.1086/160617}, \href
  {https://ui.adsabs.harvard.edu/abs/1983ApJ...264..485D} {264, 485}

\bibitem[\protect\citeauthoryear{{Galli} \& {Shu}}{{Galli} \&
  {Shu}}{1993}]{Galli&Shu1993}
{Galli} D.,  {Shu} F.~H.,  1993, \mn@doi [\apj] {10.1086/173306}, \href
  {https://ui.adsabs.harvard.edu/abs/1993ApJ...417..243G} {417, 243}

\bibitem[\protect\citeauthoryear{Gong \& Ostriker}{Gong \&
  Ostriker}{2011}]{Gong2011}
Gong H.,  Ostriker E.~C.,  2011, \mn@doi [The Astrophysical Journal]
  {10.1088/0004-637X/729/2/120}, 729, 120

\bibitem[\protect\citeauthoryear{{Gray}, {McKee}  \& {Klein}}{{Gray}
  et~al.}{2018}]{Gray2018}
{Gray} W.~J.,  {McKee} C.~F.,   {Klein} R.~I.,  2018, \mn@doi [\mnras]
  {10.1093/mnras/stx2406}, \href
  {https://ui.adsabs.harvard.edu/abs/2018MNRAS.473.2124G} {473, 2124}

\bibitem[\protect\citeauthoryear{{Guillet}, {Hennebelle}, {Pineau des
  For{\^e}ts}, {Marcowith}, {Commer{\c{c}}on}  \& {Marchand}}{{Guillet}
  et~al.}{2020}]{Guillet2020}
{Guillet} V.,  {Hennebelle} P.,  {Pineau des For{\^e}ts} G.,  {Marcowith} A.,
  {Commer{\c{c}}on} B.,   {Marchand} P.,  2020, \mn@doi [\aap]
  {10.1051/0004-6361/201937387}, \href
  {https://ui.adsabs.harvard.edu/abs/2020A&A...643A..17G} {643, A17}

\bibitem[\protect\citeauthoryear{{Harrison} et~al.,}{{Harrison}
  et~al.}{2021}]{Harrison2021}
{Harrison} R.~E.,  et~al., 2021, \mn@doi [\apj] {10.3847/1538-4357/abd94e},
  \href {https://ui.adsabs.harvard.edu/abs/2021ApJ...908..141H} {908, 141}

\bibitem[\protect\citeauthoryear{Hennebelle, Commerçon, Chabrier  \&
  Marchand}{Hennebelle et~al.}{2016}]{Hennebelle_2016}
Hennebelle P.,  Commerçon B.,  Chabrier G.,   Marchand P.,  2016, \mn@doi [The
  Astrophysical Journal Letters] {10.3847/2041-8205/830/1/L8}, 830, L8

\bibitem[\protect\citeauthoryear{{Hennebelle}, {Commer{\c{c}}on}, {Lee}  \&
  {Charnoz}}{{Hennebelle} et~al.}{2020}]{Hennebelle2020}
{Hennebelle} P.,  {Commer{\c{c}}on} B.,  {Lee} Y.-N.,   {Charnoz} S.,  2020,
  \mn@doi [\aap] {10.1051/0004-6361/201936714}, \href
  {https://ui.adsabs.harvard.edu/abs/2020A&A...635A..67H} {635, A67}

\bibitem[\protect\citeauthoryear{{Hsieh}, {Lai}, {Cheong}, {Ko}, {Li}  \&
  {Murillo}}{{Hsieh} et~al.}{2020}]{Hsieh2020}
{Hsieh} C.-H.,  {Lai} S.-P.,  {Cheong} P.-I.,  {Ko} C.-L.,  {Li} Z.-Y.,
  {Murillo} N.~M.,  2020, \mn@doi [\apj] {10.3847/1538-4357/ab7b69}, \href
  {https://ui.adsabs.harvard.edu/abs/2020ApJ...894...23H} {894, 23}

\bibitem[\protect\citeauthoryear{{Kolmogorov}}{{Kolmogorov}}{1941}]{Kolmogorov1941}
{Kolmogorov} A.,  1941, Akademiia Nauk SSSR Doklady, \href
  {https://ui.adsabs.harvard.edu/abs/1941DoSSR..30..301K} {30, 301}

\bibitem[\protect\citeauthoryear{Kuffmeier, Haugbølle  \& Åke
  Nordlund}{Kuffmeier et~al.}{2017}]{Kuffmeier_2017}
Kuffmeier M.,  Haugbølle T.,   Åke Nordlund 2017, \mn@doi [The Astrophysical
  Journal] {10.3847/1538-4357/aa7c64}, 846, 7

\bibitem[\protect\citeauthoryear{{Kuffmeier}, {Zhao}  \& {Caselli}}{{Kuffmeier}
  et~al.}{2020}]{Kuffmeier2020}
{Kuffmeier} M.,  {Zhao} B.,   {Caselli} P.,  2020, \mn@doi [\aap]
  {10.1051/0004-6361/201937328}, \href
  {https://ui.adsabs.harvard.edu/abs/2020A&A...639A..86K} {639, A86}

\bibitem[\protect\citeauthoryear{{Kuznetsova}, {Bae}, {Hartmann}  \& {Mac
  Low}}{{Kuznetsova} et~al.}{2022}]{Kuznetsova2022}
{Kuznetsova} A.,  {Bae} J.,  {Hartmann} L.,   {Mac Low} M.-M.,  2022, \mn@doi
  [\apj] {10.3847/1538-4357/ac54a8}, \href
  {https://ui.adsabs.harvard.edu/abs/2022ApJ...928...92K} {928, 92}

\bibitem[\protect\citeauthoryear{{Lam}, {Li}, {Chen}, {Tomida}  \&
  {Zhao}}{{Lam} et~al.}{2019}]{Lam2019}
{Lam} K.~H.,  {Li} Z.-Y.,  {Chen} C.-Y.,  {Tomida} K.,   {Zhao} B.,  2019,
  \mn@doi [\mnras] {10.1093/mnras/stz2436}, \href
  {https://ui.adsabs.harvard.edu/abs/2019MNRAS.489.5326L} {489, 5326}

\bibitem[\protect\citeauthoryear{{Lee}, {Li}  \& {Turner}}{{Lee}
  et~al.}{2020}]{Lee_Li2020}
{Lee} C.-F.,  {Li} Z.-Y.,   {Turner} N.~J.,  2020, \mn@doi [Nature Astronomy]
  {10.1038/s41550-019-0905-x}, \href
  {https://ui.adsabs.harvard.edu/abs/2020NatAs...4..142L} {4, 142}

\bibitem[\protect\citeauthoryear{{Lee}, {Charnoz}  \& {Hennebelle}}{{Lee}
  et~al.}{2021}]{Lee2021}
{Lee} Y.-N.,  {Charnoz} S.,   {Hennebelle} P.,  2021, \mn@doi [\aap]
  {10.1051/0004-6361/202038105}, \href
  {https://ui.adsabs.harvard.edu/abs/2021A&A...648A.101L} {648, A101}

\bibitem[\protect\citeauthoryear{{Lesur}}{{Lesur}}{2021}]{Lesur2021_review}
{Lesur} G.,  2021, \mn@doi [Journal of Plasma Physics]
  {10.1017/S0022377820001002}, \href
  {https://ui.adsabs.harvard.edu/abs/2021JPlPh..87a2001P} {87, 205870101}

\bibitem[\protect\citeauthoryear{{Lesur} et~al.,}{{Lesur}
  et~al.}{2023}]{Lesur2022PPVII}
{Lesur} G.,  et~al., 2023, in {Inutsuka} S.,  {Aikawa} Y.,  {Muto} T.,
  {Tomida} K.,   {Tamura} M.,  eds,  Astronomical Society of the Pacific
  Conference Series Vol. 534, Astronomical Society of the Pacific Conference
  Series. p.~465

\bibitem[\protect\citeauthoryear{{Li}, {Banerjee}, {Pudritz}, {J{\o}rgensen},
  {Shang}, {Krasnopolsky}  \& {Maury}}{{Li} et~al.}{2014a}]{Li2014PPVI}
{Li} Z.~Y.,  {Banerjee} R.,  {Pudritz} R.~E.,  {J{\o}rgensen} J.~K.,  {Shang}
  H.,  {Krasnopolsky} R.,   {Maury} A.,  2014a, in {Beuther} H.,  {Klessen}
  R.~S.,  {Dullemond} C.~P.,   {Henning} T.,  eds, Protostars and Planets VI.
  pp 173--194 (\mn@eprint {arXiv} {1401.2219}),
  \mn@doi{10.2458/azu_uapress_9780816531240-ch008}

\bibitem[\protect\citeauthoryear{{Li}, {Krasnopolsky}, {Shang}  \& {Zhao}}{{Li}
  et~al.}{2014b}]{Li2014}
{Li} Z.-Y.,  {Krasnopolsky} R.,  {Shang} H.,   {Zhao} B.,  2014b, \mn@doi
  [\apj] {10.1088/0004-637X/793/2/130}, \href
  {https://ui.adsabs.harvard.edu/abs/2014ApJ...793..130L} {793, 130}

\bibitem[\protect\citeauthoryear{{Marchand}, {Masson}, {Chabrier},
  {Hennebelle}, {Commer{\c{c}}on}  \& {Vaytet}}{{Marchand}
  et~al.}{2016}]{Marchand2016}
{Marchand} P.,  {Masson} J.,  {Chabrier} G.,  {Hennebelle} P.,
  {Commer{\c{c}}on} B.,   {Vaytet} N.,  2016, \mn@doi [\aap]
  {10.1051/0004-6361/201526780}, \href
  {https://ui.adsabs.harvard.edu/abs/2016A&A...592A..18M} {592, A18}

\bibitem[\protect\citeauthoryear{{Masson}, {Chabrier}, {Hennebelle}, {Vaytet}
  \& {Commer{\c{c}}on}}{{Masson} et~al.}{2016}]{Masson2016}
{Masson} J.,  {Chabrier} G.,  {Hennebelle} P.,  {Vaytet} N.,
  {Commer{\c{c}}on} B.,  2016, \mn@doi [\aap] {10.1051/0004-6361/201526371},
  \href {https://ui.adsabs.harvard.edu/abs/2016A&A...587A..32M} {587, A32}

\bibitem[\protect\citeauthoryear{{Matsumoto}, {Machida}  \&
  {Inutsuka}}{{Matsumoto} et~al.}{2017}]{Matsumoto2017}
{Matsumoto} T.,  {Machida} M.~N.,   {Inutsuka} S.-i.,  2017, \mn@doi [\apj]
  {10.3847/1538-4357/aa6a1c}, \href
  {https://ui.adsabs.harvard.edu/abs/2017ApJ...839...69M} {839, 69}

\bibitem[\protect\citeauthoryear{{Mazzei}, {Cleeves}  \& {Li}}{{Mazzei}
  et~al.}{2020}]{Mazzei2020}
{Mazzei} R.,  {Cleeves} L.~I.,   {Li} Z.-Y.,  2020, \mn@doi [\apj]
  {10.3847/1538-4357/abb67a}, \href
  {https://ui.adsabs.harvard.edu/abs/2020ApJ...903...20M} {903, 20}

\bibitem[\protect\citeauthoryear{{Mestel} \& {Spitzer}}{{Mestel} \&
  {Spitzer}}{1956}]{Mestel1956}
{Mestel} L.,  {Spitzer} L. J.,  1956, \mn@doi [\mnras]
  {10.1093/mnras/116.5.503}, \href
  {https://ui.adsabs.harvard.edu/abs/1956MNRAS.116..503M} {116, 503}

\bibitem[\protect\citeauthoryear{{Mignon-Risse}, {Gonz{\'a}lez},
  {Commer{\c{c}}on}  \& {Rosdahl}}{{Mignon-Risse}
  et~al.}{2021}]{Mignon-Risse2021}
{Mignon-Risse} R.,  {Gonz{\'a}lez} M.,  {Commer{\c{c}}on} B.,   {Rosdahl} J.,
  2021, \mn@doi [\aap] {10.1051/0004-6361/202140617}, \href
  {https://ui.adsabs.harvard.edu/abs/2021A&A...652A..69M} {652, A69}

\bibitem[\protect\citeauthoryear{{Murillo}, {van Dishoeck}, {Hacar}, {Harsono}
  \& {J{\o}rgensen}}{{Murillo} et~al.}{2022}]{Murillo2022}
{Murillo} N.~M.,  {van Dishoeck} E.~F.,  {Hacar} A.,  {Harsono} D.,
  {J{\o}rgensen} J.~K.,  2022, \mn@doi [\aap] {10.1051/0004-6361/202141250},
  \href {https://ui.adsabs.harvard.edu/abs/2022A&A...658A..53M} {658, A53}

\bibitem[\protect\citeauthoryear{{Offner}, {Clark}, {Hennebelle}, {Bastian},
  {Bate}, {Hopkins}, {Moraux}  \& {Whitworth}}{{Offner}
  et~al.}{2014}]{Offner2014PPVI}
{Offner} S.~S.~R.,  {Clark} P.~C.,  {Hennebelle} P.,  {Bastian} N.,  {Bate}
  M.~R.,  {Hopkins} P.~F.,  {Moraux} E.,   {Whitworth} A.~P.,  2014, in
  {Beuther} H.,  {Klessen} R.~S.,  {Dullemond} C.~P.,   {Henning} T.,  eds,
  Protostars and Planets VI. pp 53--75 (\mn@eprint {arXiv} {1312.5326}),
  \mn@doi{10.2458/azu_uapress_9780816531240-ch003}

\bibitem[\protect\citeauthoryear{{Offner}, {Gaches}  \& {Holdship}}{{Offner}
  et~al.}{2019}]{Offner2019}
{Offner} S. S.~R.,  {Gaches} B. A.~L.,   {Holdship} J.~R.,  2019, \mn@doi
  [\apj] {10.3847/1538-4357/ab3e02}, \href
  {https://ui.adsabs.harvard.edu/abs/2019ApJ...883..121O} {883, 121}

\bibitem[\protect\citeauthoryear{{Ohashi} et~al.,}{{Ohashi}
  et~al.}{2014}]{Ohashi2014}
{Ohashi} N.,  et~al., 2014, \mn@doi [\apj] {10.1088/0004-637X/796/2/131}, \href
  {https://ui.adsabs.harvard.edu/abs/2014ApJ...796..131O} {796, 131}

\bibitem[\protect\citeauthoryear{{Ohashi} et~al.,}{{Ohashi}
  et~al.}{2023}]{Ohashi2023}
{Ohashi} N.,  et~al., 2023, \mn@doi [\apj] {10.3847/1538-4357/acd384}, \href
  {https://ui.adsabs.harvard.edu/abs/2023ApJ...951....8O} {951, 8}

\bibitem[\protect\citeauthoryear{{Ormel}, {Paszun}, {Dominik}  \&
  {Tielens}}{{Ormel} et~al.}{2009}]{Ormel2009}
{Ormel} C.~W.,  {Paszun} D.,  {Dominik} C.,   {Tielens} A.~G.~G.~M.,  2009,
  \mn@doi [\aap] {10.1051/0004-6361/200811158}, \href
  {https://ui.adsabs.harvard.edu/abs/2009A&A...502..845O} {502, 845}

\bibitem[\protect\citeauthoryear{{Padovani}, {Ivlev}, {Galli}  \&
  {Caselli}}{{Padovani} et~al.}{2018}]{Padovani2018}
{Padovani} M.,  {Ivlev} A.~V.,  {Galli} D.,   {Caselli} P.,  2018, \mn@doi
  [\aap] {10.1051/0004-6361/201732202}, \href
  {https://ui.adsabs.harvard.edu/abs/2018A&A...614A.111P} {614, A111}

\bibitem[\protect\citeauthoryear{{Pattle}, {Fissel}, {Tahani}, {Liu}  \&
  {Ntormousi}}{{Pattle} et~al.}{2023}]{Pattle2022}
{Pattle} K.,  {Fissel} L.,  {Tahani} M.,  {Liu} T.,   {Ntormousi} E.,  2023, in
  {Inutsuka} S.,  {Aikawa} Y.,  {Muto} T.,  {Tomida} K.,   {Tamura} M.,  eds,
  Astronomical Society of the Pacific Conference Series Vol. 534, Astronomical
  Society of the Pacific Conference Series. p.~193

\bibitem[\protect\citeauthoryear{{Pineda}, {Segura-Cox}, {Caselli},
  {Cunningham}, {Zhao}, {Schmiedeke}, {Maureira}  \& {Neri}}{{Pineda}
  et~al.}{2020}]{Pineda2020}
{Pineda} J.~E.,  {Segura-Cox} D.,  {Caselli} P.,  {Cunningham} N.,  {Zhao} B.,
  {Schmiedeke} A.,  {Maureira} M.~J.,   {Neri} R.,  2020, \mn@doi [Nature
  Astronomy] {10.1038/s41550-020-1150-z}, \href
  {https://ui.adsabs.harvard.edu/abs/2020NatAs...4.1158P} {4, 1158}

\bibitem[\protect\citeauthoryear{{Sanhueza} et~al.,}{{Sanhueza}
  et~al.}{2021}]{Sanhueza2021}
{Sanhueza} P.,  et~al., 2021, \mn@doi [\apjl] {10.3847/2041-8213/ac081c}, \href
  {https://ui.adsabs.harvard.edu/abs/2021ApJ...915L..10S} {915, L10}

\bibitem[\protect\citeauthoryear{{Santos-Lima}, {de Gouveia Dal Pino}  \&
  {Lazarian}}{{Santos-Lima} et~al.}{2012}]{Santos-Lima2012}
{Santos-Lima} R.,  {de Gouveia Dal Pino} E.~M.,   {Lazarian} A.,  2012, \mn@doi
  [\apj] {10.1088/0004-637X/747/1/21}, \href
  {https://ui.adsabs.harvard.edu/abs/2012ApJ...747...21S} {747, 21}

\bibitem[\protect\citeauthoryear{Seifried, Banerjee, Pudritz  \&
  Klessen}{Seifried et~al.}{2013}]{Seifried2013}
Seifried D.,  Banerjee R.,  Pudritz R.~E.,   Klessen R.~S.,  2013, \mn@doi
  [Monthly Notices of the Royal Astronomical Society] {10.1093/mnras/stt682},
  432, 3320

\bibitem[\protect\citeauthoryear{{Seifried}, {Banerjee}, {Pudritz}  \&
  {Klessen}}{{Seifried} et~al.}{2015}]{Seifried2015_turb}
{Seifried} D.,  {Banerjee} R.,  {Pudritz} R.~E.,   {Klessen} R.~S.,  2015,
  \mn@doi [\mnras] {10.1093/mnras/stu2282}, \href
  {https://ui.adsabs.harvard.edu/abs/2015MNRAS.446.2776S} {446, 2776}

\bibitem[\protect\citeauthoryear{Shu}{Shu}{1991}]{Shu1992}
Shu F.,  1991, The Physics of Astrophysics: Gas dynamics.
Series of books in astronomy, University Science Books, \url
  {https://books.google.com/books?id=50VYSc56URUC}

\bibitem[\protect\citeauthoryear{{Stone}, {Gardiner}, {Teuben}, {Hawley}  \&
  {Simon}}{{Stone} et~al.}{2008}]{Stone2008}
{Stone} J.~M.,  {Gardiner} T.~A.,  {Teuben} P.,  {Hawley} J.~F.,   {Simon}
  J.~B.,  2008, \mn@doi [\apjs] {10.1086/588755}, \href
  {https://ui.adsabs.harvard.edu/abs/2008ApJS..178..137S} {178, 137}

\bibitem[\protect\citeauthoryear{{Stone}, {Tomida}, {White}  \&
  {Felker}}{{Stone} et~al.}{2020}]{Stone2020}
{Stone} J.~M.,  {Tomida} K.,  {White} C.~J.,   {Felker} K.~G.,  2020, \mn@doi
  [\apjs] {10.3847/1538-4365/ab929b}, \href
  {https://ui.adsabs.harvard.edu/abs/2020ApJS..249....4S} {249, 4}

\bibitem[\protect\citeauthoryear{{Thieme} et~al.,}{{Thieme}
  et~al.}{2022}]{Thieme2022}
{Thieme} T.~J.,  et~al., 2022, \mn@doi [\apj] {10.3847/1538-4357/ac382b}, \href
  {https://ui.adsabs.harvard.edu/abs/2022ApJ...925...32T} {925, 32}

\bibitem[\protect\citeauthoryear{{Tomida} \& {Stone}}{{Tomida} \&
  {Stone}}{2023}]{Tomida2023}
{Tomida} K.,  {Stone} J.~M.,  2023, \mn@doi [\apjs] {10.3847/1538-4365/acc2c0},
  \href {https://ui.adsabs.harvard.edu/abs/2023ApJS..266....7T} {266, 7}

\bibitem[\protect\citeauthoryear{{Tomida}, {Tomisaka}, {Matsumoto}, {Hori},
  {Okuzumi}, {Machida}  \& {Saigo}}{{Tomida} et~al.}{2013}]{Tomida2013}
{Tomida} K.,  {Tomisaka} K.,  {Matsumoto} T.,  {Hori} Y.,  {Okuzumi} S.,
  {Machida} M.~N.,   {Saigo} K.,  2013, \mn@doi [\apj]
  {10.1088/0004-637X/763/1/6}, \href
  {https://ui.adsabs.harvard.edu/abs/2013ApJ...763....6T} {763, 6}

\bibitem[\protect\citeauthoryear{{Tsukamoto}, {Machida}  \&
  {Inutsuka}}{{Tsukamoto} et~al.}{2023a}]{Tsukamoto2023}
{Tsukamoto} Y.,  {Machida} M.~N.,   {Inutsuka} S.-i.,  2023a, \mn@doi [arXiv
  e-prints] {10.48550/arXiv.2303.10419}, \href
  {https://ui.adsabs.harvard.edu/abs/2023arXiv230310419T} {p. arXiv:2303.10419}

\bibitem[\protect\citeauthoryear{{Tsukamoto} et~al.,}{{Tsukamoto}
  et~al.}{2023b}]{Tsukamoto2022PPVII}
{Tsukamoto} Y.,  et~al., 2023b, in {Inutsuka} S.,  {Aikawa} Y.,  {Muto} T.,
  {Tomida} K.,   {Tamura} M.,  eds,  Astronomical Society of the Pacific
  Conference Series Vol. 534, Astronomical Society of the Pacific Conference
  Series. p.~317

\bibitem[\protect\citeauthoryear{{Tu}, {Li}  \& {Lam}}{{Tu}
  et~al.}{2022}]{Tu2022}
{Tu} Y.,  {Li} Z.-Y.,   {Lam} K.~H.,  2022, \mn@doi [\mnras]
  {10.1093/mnras/stac2030}, \href
  {https://ui.adsabs.harvard.edu/abs/2022MNRAS.515.4780T} {515, 4780}

\bibitem[\protect\citeauthoryear{{Valdivia-Mena} et~al.,}{{Valdivia-Mena}
  et~al.}{2022}]{Valdivia-Mena2022}
{Valdivia-Mena} M.~T.,  et~al., 2022, \mn@doi [\aap]
  {10.1051/0004-6361/202243310}, \href
  {https://ui.adsabs.harvard.edu/abs/2022A&A...667A..12V} {667, A12}

\bibitem[\protect\citeauthoryear{{Vlemmings} et~al.,}{{Vlemmings}
  et~al.}{2019}]{Vlemmings2019}
{Vlemmings} W.~H.~T.,  et~al., 2019, \mn@doi [\aap]
  {10.1051/0004-6361/201935459}, \href
  {https://ui.adsabs.harvard.edu/abs/2019A&A...624L...7V} {624, L7}

\bibitem[\protect\citeauthoryear{{Wurster} \& {Lewis}}{{Wurster} \&
  {Lewis}}{2020a}]{Wurster2020_I}
{Wurster} J.,  {Lewis} B.~T.,  2020a, \mn@doi [\mnras]
  {10.1093/mnras/staa1339}, \href
  {https://ui.adsabs.harvard.edu/abs/2020MNRAS.495.3795W} {495, 3795}

\bibitem[\protect\citeauthoryear{{Wurster} \& {Lewis}}{{Wurster} \&
  {Lewis}}{2020b}]{Wurster2020_II}
{Wurster} J.,  {Lewis} B.~T.,  2020b, \mn@doi [\mnras]
  {10.1093/mnras/staa1340}, \href
  {https://ui.adsabs.harvard.edu/abs/2020MNRAS.495.3807W} {495, 3807}

\bibitem[\protect\citeauthoryear{{Wurster} \& {Li}}{{Wurster} \&
  {Li}}{2018}]{Wurster2018}
{Wurster} J.,  {Li} Z.-Y.,  2018, \mn@doi [Frontiers in Astronomy and Space
  Sciences] {10.3389/fspas.2018.00039}, \href
  {https://ui.adsabs.harvard.edu/abs/2018FrASS...5...39W} {5, 39}

\bibitem[\protect\citeauthoryear{{Xu} \& {Kunz}}{{Xu} \&
  {Kunz}}{2021a}]{Xu2021}
{Xu} W.,  {Kunz} M.~W.,  2021a, \mn@doi [\mnras] {10.1093/mnras/stab314}, \href
  {https://ui.adsabs.harvard.edu/abs/2021MNRAS.502.4911X} {502, 4911}

\bibitem[\protect\citeauthoryear{{Xu} \& {Kunz}}{{Xu} \&
  {Kunz}}{2021b}]{XuKunz2021}
{Xu} W.,  {Kunz} M.~W.,  2021b, \mn@doi [\mnras] {10.1093/mnras/stab2715},
  \href {https://ui.adsabs.harvard.edu/abs/2021MNRAS.508.2142X} {508, 2142}

\bibitem[\protect\citeauthoryear{{Yen} et~al.,}{{Yen} et~al.}{2014}]{Yen2014}
{Yen} H.-W.,  et~al., 2014, \mn@doi [\apj] {10.1088/0004-637X/793/1/1}, \href
  {https://ui.adsabs.harvard.edu/abs/2014ApJ...793....1Y} {793, 1}

\bibitem[\protect\citeauthoryear{{Yen}, {Takakuwa}, {Gu}, {Hirano}, {Lee},
  {Liu}, {Liu}  \& {Wu}}{{Yen} et~al.}{2019}]{Yen2019}
{Yen} H.-W.,  {Takakuwa} S.,  {Gu} P.-G.,  {Hirano} N.,  {Lee} C.-F.,  {Liu}
  H.~B.,  {Liu} S.-Y.,   {Wu} C.-J.,  2019, \mn@doi [\aap]
  {10.1051/0004-6361/201834209}, \href
  {https://ui.adsabs.harvard.edu/abs/2019A&A...623A..96Y} {623, A96}

\bibitem[\protect\citeauthoryear{{Zhao}, {Li}, {Nakamura}, {Krasnopolsky}  \&
  {Shang}}{{Zhao} et~al.}{2011}]{Zhao2011}
{Zhao} B.,  {Li} Z.-Y.,  {Nakamura} F.,  {Krasnopolsky} R.,   {Shang} H.,
  2011, \mn@doi [\apj] {10.1088/0004-637X/742/1/10}, \href
  {https://ui.adsabs.harvard.edu/abs/2011ApJ...742...10Z} {742, 10}

\bibitem[\protect\citeauthoryear{{Zhao} et~al.,}{{Zhao}
  et~al.}{2020}]{ZhaoTomida2020}
{Zhao} B.,  et~al., 2020, \mn@doi [\ssr] {10.1007/s11214-020-00664-z}, \href
  {https://ui.adsabs.harvard.edu/abs/2020SSRv..216...43Z} {216, 43}

\makeatother
\end{thebibliography}








\bsp	
\label{lastpage}
\end{document}